      [1999/12/01 v1.4c Il Nuovo Cimento]
\documentclass{cimento}

\usepackage{graphicx}  
\title{Hot Gas in Galaxy Clusters: Theory and Simulations}
\author{Michael L. Norman\from{ins:x}}
\instlist{\inst{ins:x} Physics Department and CASS, UC San Diego, USA}
\begin{document}

\maketitle

\begin{abstract}
We review the theory of the formation of galaxy clusters and discuss their 
role as cosmological probes. We begin with the standard cosmological 
framework where we discuss the origin of the CDM matter power spectrum
and the growth of density fluctuations in the linear regime. We then summarize 
the spherical top-hat model for the nonlinear growth of fluctuations from which 
scaling relations and halo statistics are derived. 
Numerical methods for simulating gas in galaxy 
clusters are then overviewed with an emphasis on multiscale hydrodynamic 
simulations of cluster ensembles. Results of hydrodynamic AMR simulations 
are described which compare cluster internal and statistical properties as a 
function of their assumed baryonic processes. Finally, we compare various 
methods of measuring cluster masses using X-ray and the thermal 
Sunyaev-Zeldovich effect (SZE). We find that SZE offers great promise for 
precision measurements in raw samples of high-z clusters. 
\end{abstract}

\section{Introduction}
The Sunyaev-Zeldovich Effect (SZE) detectable in galaxy clusters has emerged 
as a powerful new probe of the low to intermediate redshift universe (see 
articles by Birkinshaw {\&} Rephaeli in this volume, as well the review by 
Carlstrom, et al. \cite{carlstrom02}. Within the prevailing theory of cosmological 
structure formation, galaxy clusters form in rare, massive peaks of the 
cosmic density field. Because of natural biasing, such regions get a ``head 
start'' on structure formation on all scales smaller than the cluster scale. 
As a consequence, galaxy clusters at the present epoch contain the oldest 
objects in the universe in an evolutionary sense \cite{springel05}. 
This makes galaxy clusters intrinsically interesting as astrophysical 
objects, worthy of study observationally, theoretically, and 
computationally. 

However, much of the current interest stems from the potential use of galaxy 
clusters as cosmological probes. As discussed in more detail below, the 
space density of galaxy clusters as a function of cosmological redshift is 
sensitive to the RMS mass fluctuations on scales of 10$^{14-15} M_{\odot}$, 
which depends on $\Omega _{m}$, the mean mass density of the universe, and 
to a lesser extent, $\Omega _{de}$, the dark energy density of the 
universe. Attempts to deduce $\Omega _{m}$ based on X-ray surveys have met 
with some success \cite{Rosati02}, but they have been 
hampered by the fact that at these wavebands cluster samples become sparse 
at z$>$1 owing to their low surface brightness. Because the SZE 
is intrinsically redshift independent, one has the possibility of detecting 
clusters over a wide range of redshifts. Blind surveys with sufficient 
sensitivity can in principle detect clusters from z=0 to their formation 
redshift $z \leq 1.5$ \cite{carlstrom02}, paving the way for more precise 
cosmological parameter measurements. Follow-up pointed observations of a 
large sample of galaxy clusters over a range of redshifts would enable a 
detailed study of their formation and evolution. Such studies would confirm 
or modify our theory of structure formation, improve our understanding of 
galaxy evolution, and reveal a great deal about the complex physical 
processes operating in the intracluster medium (ICM). 

This paper summarizes four lectures the author delivered at the Varenna 
Summer School entitled ``Background Microwave Radiation and Intracluster
Cosmology'', 
held July 2004 in Varenna, Italy. Originally, the organizers asked me to 
deliver three lectures covering numerical simulations of galaxy clusters, as 
well as to review the basics of cosmological structure formation, of which 
galaxy clusters are just one aspect. The first lecture of the school was to 
have been given by Dr. Rocky Kolb on the cosmological standard model and the 
linear growth of density perturbations. When he was unable to attend the 
school, that responsibility fell to me, increasing my task to four lectures. 
Fortunately, Dr. Kolb's lecture slides were made available to me, which I 
used verbatim. The following Section 2 follows closely the content and 
organization of Dr. Kolb's lecture notes, while Sections 3-5 are my own. 
Section 3 reviews key concepts and results from structure formation theory 
that provide the vocabulary and framework for interpreting observations and 
simulations of galaxy clusters. Section 4 discusses the technical 
challenges associated with simulating gas in galaxy clusters and reviews the 
numerical methods we have employed. Section 5 presents results of numerical 
simulations of statistical ensembles of galaxy glusters whose goal is to 
understand how observables such as X-ray luminosity, emission-weighted 
temperature, and SZE depend on cluster mass and baryonic physics. 

In line with the character of the summer school, I have attempted to be 
pedagogical, emphasizing the key concepts and results that a student needs 
to know if he/she wants to understand the current literature or do 
research in this area. Literature citations are kept to a minimum, except 
for textbooks, reviews, and research papers that I found to be particularly 
helpful in preparing this article.

\section {Cosmological framework and perturbation growth in the 
linear regime}

Our modern theory of the structure and evolution of the universe, along with 
the observational data which support it, is admirably presented in a recent 
textbook by Dodelson \cite{dodelson03}. 
Remarkable observational progress has been made in the past two 
decades which has strengthened our confidence in the correctness of the 
hot, relativistic, expanding universe model (Big Bang), has measured the 
universe's present mass-energy contents and kinematics, and lent strong 
support to the notion of a very early, inflationary phase. Moreover, 
observations of high redshift supernovae unexpectedly have revealed that the 
cosmic expansion is accelerating at the present time, implying the existence 
of a pervasive, dark energy field with negative pressure \cite{Perlmutter03}. This surprising discovery has enlivened observational efforts to 
accurately measure the cosmological parameters over as large a fraction of 
the age of the universe as possible, especially over the redshift interval 0 
$<$ z $<$ 1.5 which, according to current estimates, spans the 
deceleration-acceleration transition. These efforts include large surveys of 
galaxy large scale structure, galaxy clusters, weak lensing, the Lyman alpha 
forest, and high redshift supernovae, all of which span the relevant 
redshift range. Except for the supernovae, all other techniques rely on 
measurements of cosmological structure in order to deduce cosmological 
parameters.

\subsection{Cosmological standard model}
The dynamics of the expanding universe is described by the two Friedmann 
equations derived from Einstein's theory of general relativity under the 
assumption of homogeneity and isotropy. The expansion rate at time $t$ is 
given by
\begin{equation}\label{eq1}
H^2(t)\equiv \left( {\frac{\dot {a}}{a}} \right)^2=\frac{8\pi 
G}{3}\sum\limits_i {\rho _i } -\frac{k}{a^2}+\frac{\Lambda }{3}
\end{equation}
where $H(t)$ is the Hubble parameter and $a(t)$ is the FRW scale factor at time 
$t$. The first term on the RHS is proportional to the sum over 
all energy densities in 
the universe $\rho _{i }$ including baryons, photons, neutrinos, dark 
matter and dark energy. We have explicitly pulled the dark energy term out 
of the sum and placed it in the third term assuming it is a constant (the 
cosmological constant). The second term is the curvature term, where 
$k=0,\pm 1$ for zero, positive, negative curvature, respectively. Equation (\ref{eq1}) 
can be cast in a form useful for numerical integration if we introduce 
$\Omega $ parameters:
\begin{equation}\label{eq2}
\Omega _i \equiv \frac{8\pi G}{3H^2}\rho _i ,\mbox{ }\Omega _\Lambda \equiv 
\frac{8\pi G}{3H^2}\rho _\Lambda =\frac{\Lambda }{3H^2},\mbox{ }\Omega_k 
\equiv \frac{-k}{(aH)^2}
\end{equation}
Dividing equation (\ref{eq1}) by $H^2$ we get the sum rule 1=$\Omega 
_{m}+\Omega _{k}+\Omega _{\Lambda }$, which is true at all times, 
where $\Omega _{m}$ is the sum over all $\Omega _{i}$ excluding dark 
energy. At the present time $H(t)=H_{0}, a=1$, and cosmological density 
parameters become
\begin{equation}\label{eq3}
\Omega _i (0)=\frac{8\pi G}{3H_0^2 }\rho _i (0),\mbox{ }\Omega _\Lambda 
(0)=\frac{\Lambda }{3H_0^2 },\mbox{ }\Omega _k (0)=\frac{-k}{H_0^2 }
\end{equation}
Equation (\ref{eq1}) can then be manipulated into the form
\begin{equation}\label{eq4}
\dot {a}=H_0 [\Omega _m (0)(a^{-1}-1)+\Omega _\gamma (0)(a^{-2}-1)+\Omega 
_\Lambda (0)(a^2-1)+1]^{1/2}
\end{equation}
Here we have explicitly introduced a density parameter for the background 
radiation field $\Omega _{\gamma }$ and used the fact that matter and radiation 
densities scale as a$^{-3}$ and a$^{-4}$, respectively,
and we have used the sum rule to eliminate $\Omega _{k}$. Equation (\ref{eq4}) is 
equation (\ref{eq1}) expressed in terms of the \textit{current} values of the density and Hubble 
parameters, and makes explicit the scale factor dependence of the various 
contributions to the expansion rate. In particular, it is clear that the 
expansion rate is dominated first by radiation, then by matter, and finally 
by the cosmological constant.

Current measurements of the cosmological parameters by different techniques 
\cite{spergel03} yield the following numbers [(0) notation 
suppressed]:
\[
\begin{array}{l}
 h\equiv H_0 /(100km/s/Mpc)\approx 0.72 \\ 
 \Omega _{total} \approx 1,\mbox{ }\Omega _\Lambda \approx 0.73\mbox{, 
}\Omega _m =\Omega _{cdm} +\Omega _b \approx 0.27,\Omega _k \approx 0 \\ 
 \Omega _b \approx 0.04,\mbox{ }\Omega _\nu \approx 0.005,\mbox{ }\Omega 
_\gamma \approx 0.00005 \\ 
 \end{array}
\]
This set of parameters is referred to as the concordance model \cite{bops99}, and describes a spatially flat, low matter density, high dark 
energy density universe in which baryons, neutrinos, and photons make a 
negligible contribution to the large scale dynamics. Most of the matter in 
the universe is cold dark matter (CDM) whose dynamics is discussed below. As 
we will also see below, baryons and photons make an important contribution 
to shaping of the matter power spectrum despite their small contribution to
the present-day energy budget. Understanding the evolution of 
baryons in nonlinear structure formation is essential to interpret X-ray and 
SZE observations of galaxy clusters. 

The second Friedmann equation relates the second time derivative of the 
scale factor to the cosmic pressure $p $ and energy density\textit{ $\rho $}
\begin{equation}\label{eq5}
\frac{\ddot {a}}{a}=-\frac{4\pi G}{3}(\rho +3p),\mbox{ }\rho =\sum\limits_i 
{\rho _i } =\rho _m +\rho _\gamma +\rho _\Lambda 
\end{equation}
$p$ and $\rho $ are related by an equation of state $p_{i}=w_{i}\rho 
_{i}$, with $w_{m}$=0, $w_{\gamma }$=1/3, and $w_{\Lambda }= -1$. We thus 
have
\begin{equation}\label{eq6}
\frac{\ddot {a}}{a}=-\frac{4\pi G}{3}(\rho _m +2\rho _\gamma -2\rho _\Lambda 
)\mbox{. }
\end{equation}
Expressed in terms of the current values for the cosmological parameters we 
have
\begin{equation}\label{eq7}
\frac{\ddot {a}}{a}=-\frac{1}{2}H_0^2 [\Omega _m (0)a^{-3}+2\Omega _\gamma 
(0)a^{-4}-2\Omega _\Lambda (0)]\mbox{. }
\end{equation}
Evaluating equation \ref{eq7} using the concordance parameters, we see the 
universe is currently accelerating $\ddot {a}\approx 0.6H_0^2 \mbox{ }$ . 
Assuming the dark energy density is a constant, the acceleration began when
\begin{equation}\label{eq8}
a\equiv \frac{1}{1+z}=\left( {\frac{\Omega _m (0)}{2\Omega _\Lambda (0)}} 
\right)^{\mbox{1/3}}\mbox{ }\approx 0.57
\end{equation}
or $z\sim 0.75$.

\subsection{The Linear power spectrum}

\begin{figure}[htbp]
\centerline{\includegraphics[width=4.15in,height=3.8in]{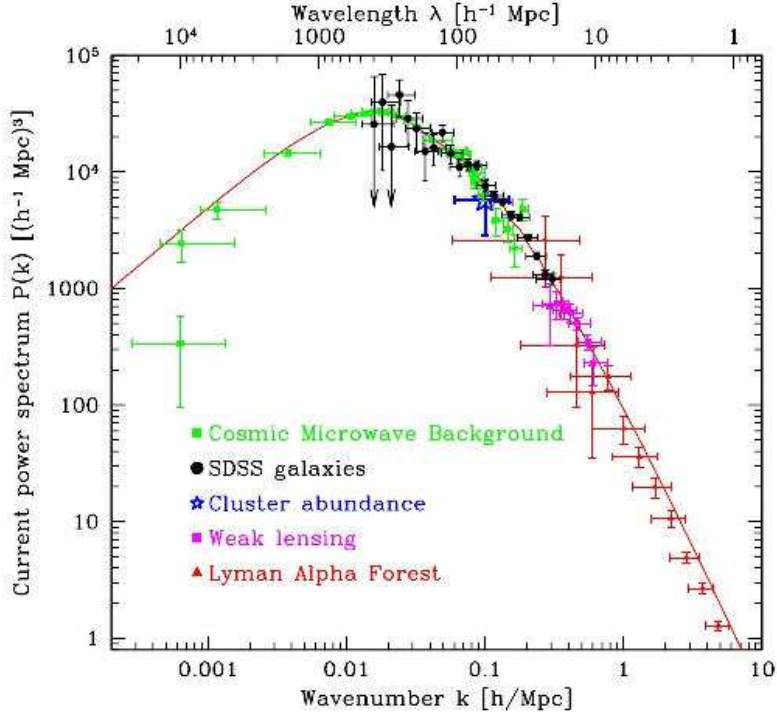}}
\caption{Linear matter power spectrum P(k) versus wavenumber extrapolated 
to z=0, from various measurements of cosmological structure. The best fit 
$\Lambda $CDM model is shown as a solid line. From \cite{tegmark03}.}
\label{fig1}
\end{figure}

Cosmic structure results from the amplification of primordial density 
fluctuations by gravitational instability. The power spectrum of matter 
density fluctuations has now been measured with considerable accuracy across 
roughly four decades in scale. Figure \ref{fig1} shows the latest results, 
taken from reference
\cite{tegmark03}. Combined in this figure are measurements using cosmic 
microwave background (CMB) anisotropies, galaxy large scale structure, weak 
lensing of galaxy shapes, and the Lyman alpha forest, in order of decreasing 
comoving wavelength. In addition, there is a single data point for galaxy 
clusters, whose current space density measures the amplitude of the power 
spectrum on 8 h$^{-1}$ Mpc scales \cite{wef93}. 
Superimposed on the data is the predicted $\Lambda $CDM linear power 
spectrum at z=0 for the concordance model parameters. As one can see, the 
fit is quite good. In actuality, the concordance model parameters are 
determined by fitting the data. A rather complex statistical machinery 
underlies the determination of cosmological parameters, and is discussed in 
Dodelson (2003, Ch. 11). The fact that modern CMB and LSS data agree over a 
substantial region of overlap gives us confidence in the correctness of the 
concordance model. In this section, we define the power spectrum 
mathematically, and review the basic physics which determines its shape. 
Readers wishing a more in depth treatment are referred to references 
\cite{dodelson03,kolbturner90}.

At any epoch $t$ (or $a$ or $z)$ express the matter density in the universe in terms 
of a mean density and a local fluctuation:
\begin{equation}\label{eq9}
\rho (\vec {x})=\bar {\rho }(1+\delta (\vec {x}))
\end{equation}
where $\delta (\vec {x})$is the density contrast. Expand $\delta (\vec 
{x})$ in Fourier modes:
\begin{equation}\label{eq10}
\delta (\vec {x})\equiv \frac{\rho (\vec {x})-\bar {\rho }}{\bar {\rho 
}}=\int {\delta (\vec {k})\exp (-i\vec {k}\cdot \vec {x})d^3} k.
\end{equation}
The autocorrelation function of $\delta (\vec {x})$ defines the power 
spectrum through the relations
\begin{equation}\label{eq11}
\left\langle {\delta (\vec {x})\delta (\vec {x})} \right\rangle 
=\int\limits_0^\infty {\frac{dk}{k}} \frac{k^3\left| {\delta ^2(\vec {k})} 
\right|}{2\pi ^2}=\int\limits_0^\infty {\frac{dk}{k}} \frac{k^3P(k)}{2\pi 
^2}=\int\limits_0^\infty {\frac{dk}{k}} \Delta ^2(k)
\end{equation}
where we have the definitions
\begin{equation}\label{eq12}
P(k)\equiv \left| {\delta ^2(\vec {k})} \right|,\mbox{ and }\Delta 
^2(k)\equiv \frac{k^3P(k)}{2\pi ^2}.
\end{equation}
The quantity $\Delta ^2(k)$ is called the dimensionless power spectrum and is 
an important function in the theory of structure formation. $\Delta 
^2(k)$ measures the contribution of perturbations per unit logarithmic 
interval at wavenumber $k$ to the variance in the matter density fluctuations. 
The $\Lambda $CDM power spectrum asymptotes to $P(k)\sim k^{1}$ for small 
$k$, and $P(k)\sim k^{-3}$ for large $k$, with a peak a $k^{\star}\sim 2\times 10^{-2}$ h 
Mpc$^{-1}$ corresponding to $\lambda^{\star}\sim $350 h$^{-1}$ Mpc. 
$\Delta ^2(k)$ is thus asymptotically flat at high $k$, but drops off as 
$k^{4}$ at small $k$. We therefore see that most of the variance
in the cosmic density field in the universe at the present epoch is on 
scales $\lambda <  \lambda^{\star}.$ 

\begin{figure}[htbp]
\centerline{\includegraphics[width=4in,height=3in]{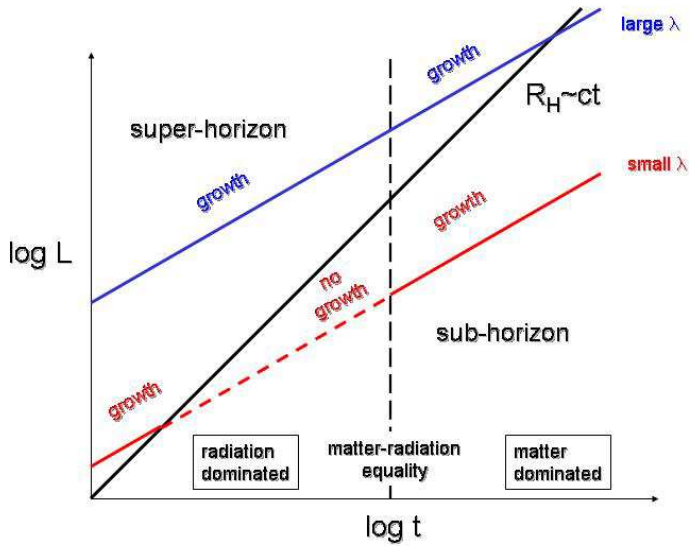}}
\caption{The tale of two fluctuations. A fluctuation which is superhorizon 
scale at matter-radiation equality grows always, while a fluctuation which 
enters the horizon during the radiation dominated era stops growing in 
amplitude until the matter dominated era begins.}
\label{fig2}
\end{figure}

What is the origin of the power spectrum shape? Here we review the basic ideas. 
Within the inflationary paradigm, it is believed that quantum mechanical 
(QM) fluctuations in the very early universe were stretched to macroscopic 
scales by the large expansion factor the universe underwent during 
inflation. Since QM fluctuations are random, the primordial density 
perturbations should be well described as a Gaussian random field. 
Measurements of the Gaussianity of the CMB anisotropies \cite{komatsu03} have confirmed 
this. The primordial power spectrum is parameterized as a power law $P_p 
(k)\propto k^n$, with $n=1$ corresponding to scale-invariant spectrum 
proposed by Harrison and Zeldovich on the grounds that any other value would 
imply a preferred mass scale for fluctuations entering the Hubble horizon. 
Large angular scale CMB anisotropies measure the primordial power spectrum 
directly since they are superhorizon scale. Observations with the WMAP 
satellite are consistent with $n=1$. 

To understand the origin of the spectrum, we need to understand how the 
amplitude of a 
fluctuation of fixed comoving wavelength $\lambda$ grows with 
time. Regardless of its wavelength, the fluctuation will pass through 
the Hubble horizon as illustrated in Fig. \ref{fig2}. This is because the Hubble 
radius grows linearly with time, while the proper wavelength a$\lambda $ 
grows more slowly with time. It is easy to show from Eq. \ref{eq1} that in 
the radiation-dominated era, $a\sim t^{1/2}$, and in the matter-dominated era 
(prior to the onset of cosmic acceleration) $a\sim t^{2/3}$. Thus, inevitably, 
a fluctuation will transition from superhorizon to subhorizon scale. We are 
interested in how the amplitude of the fluctuation evolves during these two 
phases. Here we merely state the results of perturbation theory (e.g., 
Dodelson 2003, Ch. 7).

\begin{figure}[htbp]
\begin{tabular}{c}
\centerline{\includegraphics[width=4in,height=2in]{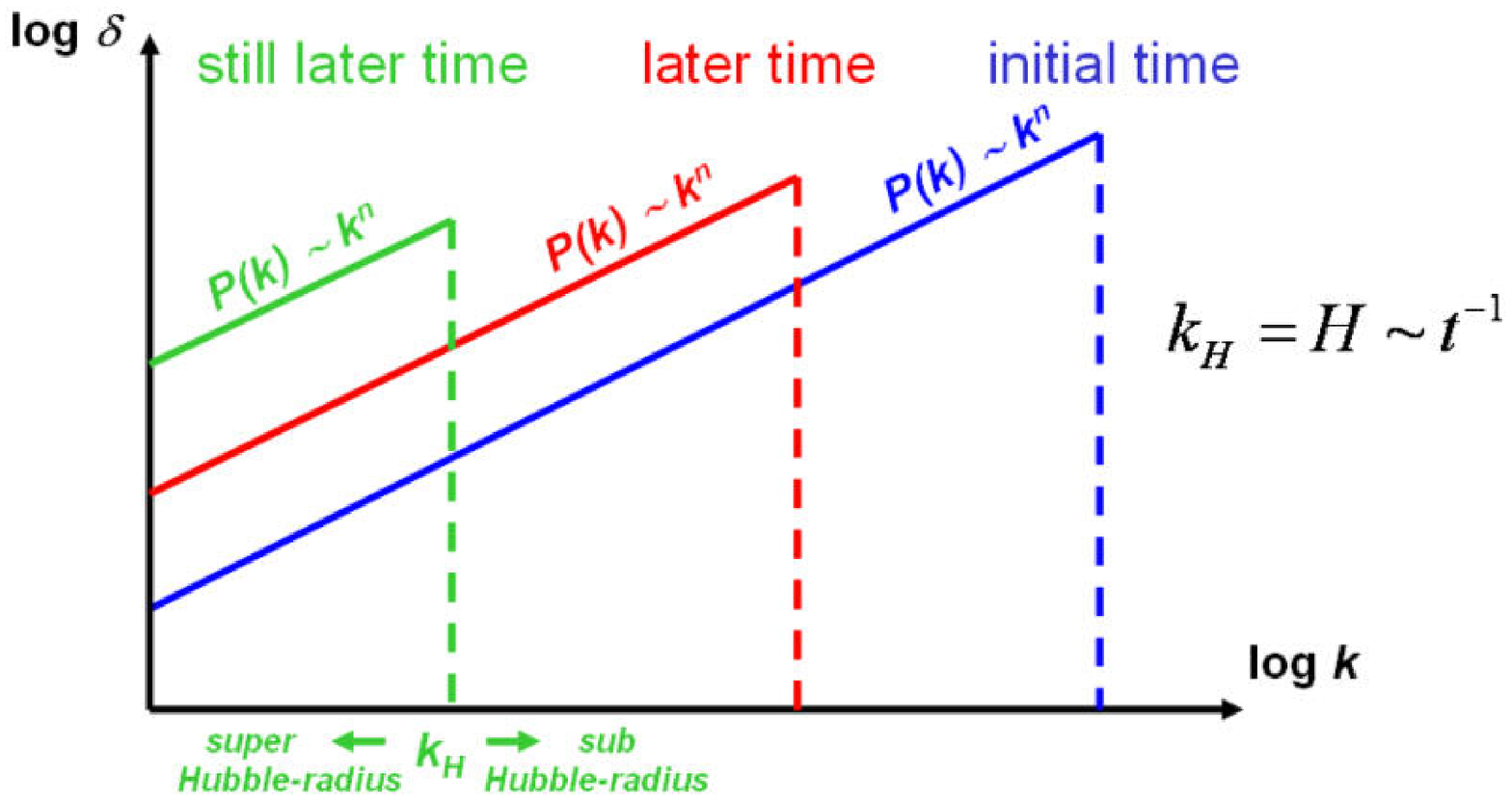}} \\
\centerline{\includegraphics[width=4.4in,height=2.2in]{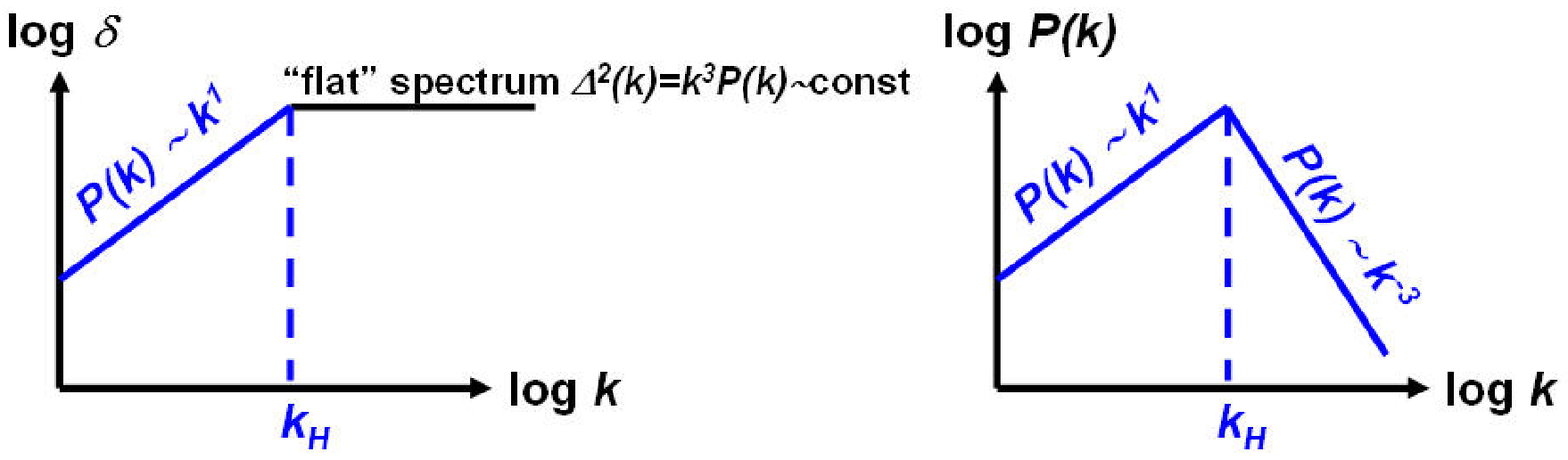}}
\end{tabular}
\caption{ a) Evolution of the primordial power spectrum on superhorizon 
scales during the radiaton dominated era. b) Scale-free spectrum produces a 
constant contribution to the density variance per logarithmic wavenumber 
interval entering the Hubble horizon (no preferred scale) c) resulting 
matter power spectrum, super- and sub-horizon. Figures courtesy Rocky Kolb.}
\label{fig3}
\end{figure}

\subsection{Growth of fluctuations in the linear regime }

To calculate the growth of superhorizon scale fluctuations requires general relativistic 
perturbation theory, while subhorizon scale perturbations can be analyzed 
using a Newtonian Jeans analysis. We are interested in scalar 
density perturbations, because these couple to the stress tensor of the 
matter-radiation field. Vector perturbations (e.g., fluid turbulence)
are not sourced by the 
stress-tensor, and decay rapidly due to cosmic expansion. Tensor 
perturbations are gravity waves, and also do not couple to the 
stress-tensor. A detailed analysis for the scalar perturbations yields the 
following results. In the \underline {radiation dominated era}, 
\[
\begin{array}{l}
 \delta _+ (t)=\delta _+ (t_i )(t/t_i )\mbox{ superhorizon scales} \\ 
 \delta _+ (t)=constant                \mbox{ ~~~subhorizon scales} \\ 
 \end{array}
\]
while in the \underline {matter dominated era}, 
\[
\begin{array}{l}
 \delta _+ (t)=\delta _+ (t_i )(t/t_i )^{2/3}\mbox{ superhorizon scales} \\ 
 \delta _+ (t)=\delta _+ (t_i )(t/t_i )^{2/3}\mbox{ subhorizon scales} \\ 
 \end{array}
\]
This is summarized in Fig. \ref{fig2}, where we consider two fluctuations of 
different comoving wavelengths, which we will call large and small. The 
large wavelength perturbation remains superhorizon through matter-radiation 
equality (MRE), and enters the horizon in the matter dominated era. Its amplitude 
will grow as $t$ in the radiation dominated era, and as $t^{2/3}$ in the matter 
dominated era. It will continue to grow as $t^{2/3}$ after it becomes subhorizon 
scale. The small wavelength perturbation becomes subhorizon before 
MRE. Its amplitude will grow as $t$ while it is 
superhorizon scale, remain constant while it is subhorizon during the 
radiation dominated era, and then grow as $t^{2/3}$ during the matter-dominated 
era. 

Armed with these results, we can understand what is meant by a scale-free 
primordial power spectrum (the Harrison-Zeldovich power spectrum.) We are 
concerned with perturbation growth in the very early universe during the 
radiation dominated era. Superhorizon scale perturbation amplitudes grow as 
$t$, and then cease to grow after they have passed through the Hubble horizon. 
We can define a Hubble wave number $k_H \equiv 2\pi /R_H \propto 
t^{-1}.$ Fig. 3a shows the primordial power spectrum at three instants in time 
for k$<$k$_{H}$. We see that the fluctuation amplitude at k=k$_{H}$(t) 
depends on primordial power spectrum slope n. The scale-free spectrum is the 
value of n such that $\Delta ^2(k_{H}(t))$=constant for k$>$k$_{H}$. A simple 
analysis shows that this implies n=1. Since $\Delta ^2(k)\propto k^3P(k)$, we 
then have
\[
\begin{array}{l}
 P(k)\propto k^1,\mbox{ }k\le k_H \\ 
 P(k)\propto k^{-3},\mbox{ }k>k_H \\ 
 \end{array}
\]

In actuality, the power spectrum has a smooth maximum, rather than a peak as 
shown in Fig. 3c. This smoothing is caused by the different rates of growth 
before and after matter-radiation equality. 
The transition from radiation to matter-dominated is not 
instantaneous. Rather, the expansion rate of the universe changes smoothly 
through equality, as given by Eq. 1, and consequently so do the temporal 
growth rates. The position of the peak of the power spectrum is sensitive to 
the when the universe reached matter-radiation equality, and hence is a 
probe of $\Omega _\gamma /\Omega _m $.

\begin{figure}[htbp]
\includegraphics[width=2.5in,height=1.7in]{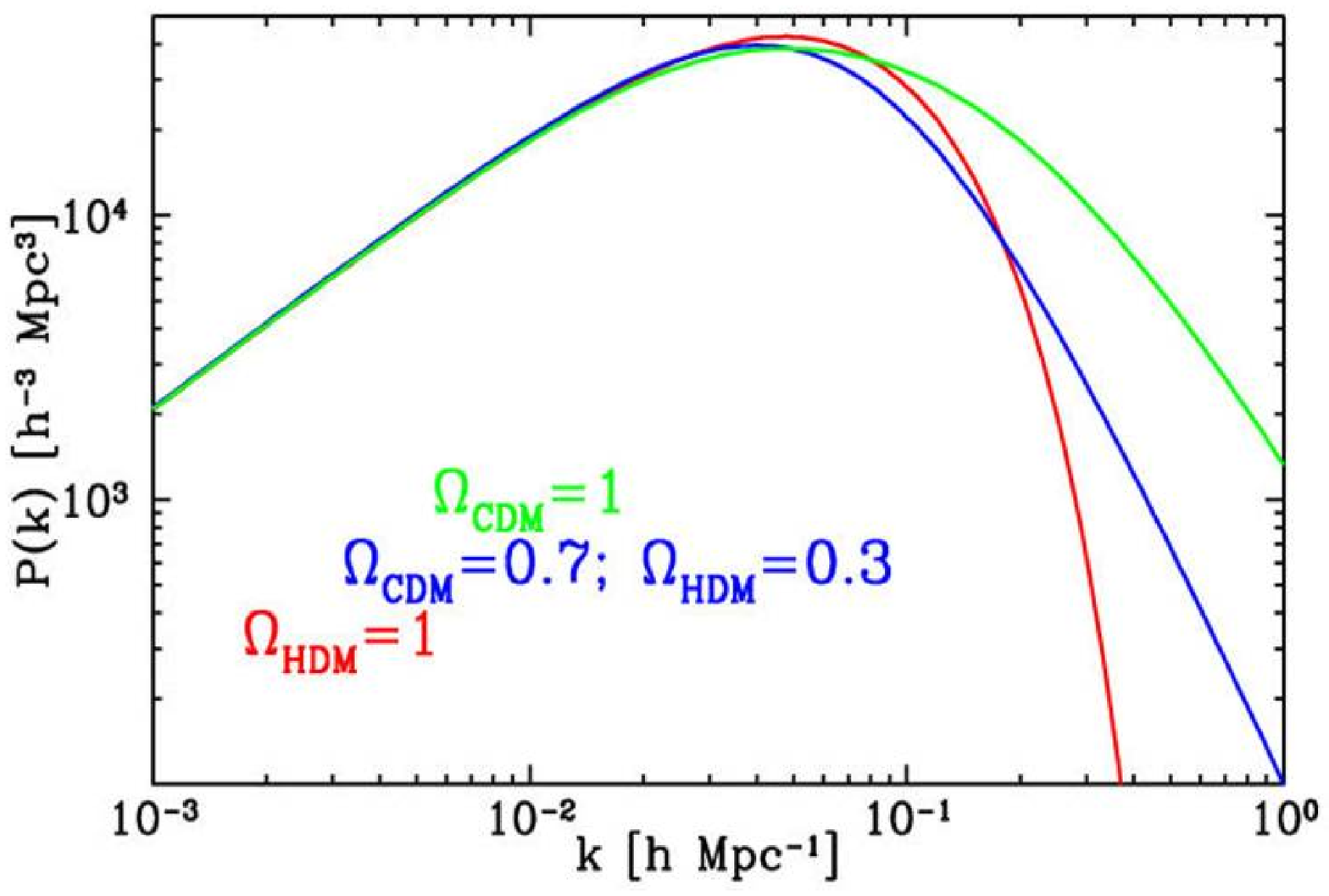}
\includegraphics[width=2.5in,height=1.7in]{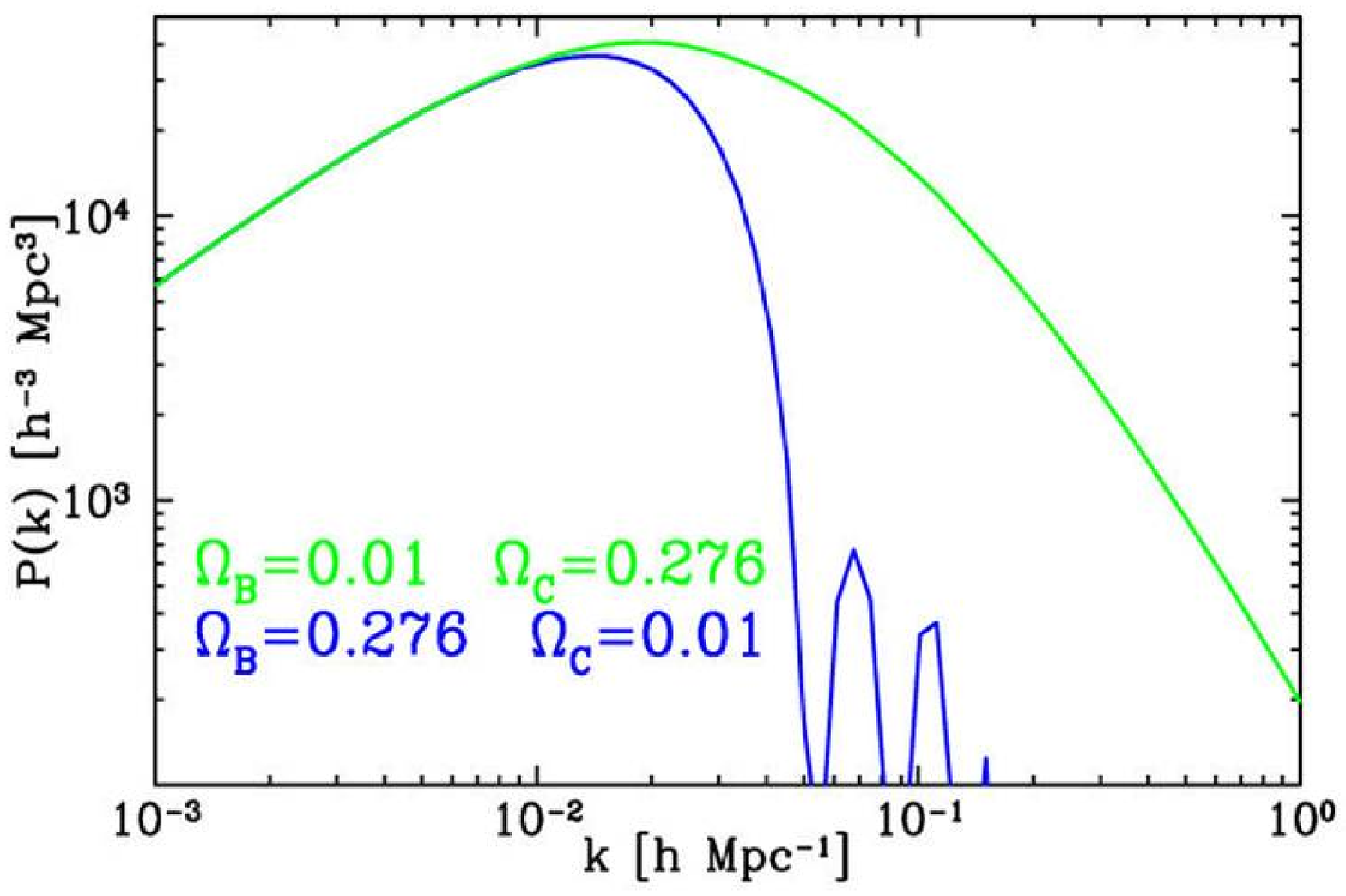}
\caption{Effect of dissipative processes on the evolved power spectrum. Left: Effect of collisionless damping (free streaming) in the dark matter. Right: Effect of collisional damping (Silk damping) in the matter-radiation fluid. Figures courtesy Rocky Kolb.}
\label{fig4}
\end{figure}

Once a fluctuation becomes sub-horizon, dissipative processes modify the 
shape of the power spectrum in a scale-dependent way. Collisionless matter 
will freely stream out of overdense regions and smooth out the 
inhomogeneities. The faster the particle, the larger its free streaming 
length. Particles which are relativistic at MRE, such as light neutrinos, 
are called hot dark matter (HDM). They have a large free-streaming length, 
and consequently damp the power spectrum over a large range of k. Weakly 
Interacting Massive Particles (WIMPs) which are nonrelativistic at MRE, are 
called cold dark matter (CDM), and modify the power spectrum very little
(Fig. \ref{fig4}). 
Baryons are tightly coupled to the radiation field by electron 
scattering prior to recombination. During rcombination, the photon mean-free 
path becomes large. As photons stream out of dense regions, they drag 
baryons along, erasing density fluctuations on small scales. This process is 
called Silk damping, and results in damped oscillations of the baryon-photon 
fluid once they become subhorizon scale. The magnitude of this effect is 
sensitive to the ratio of baryons to collisionless matter, as shown in Fig. 
\ref{fig4}. 

\section {Analytic models for nonlinear growth, virial scaling \\ 
relations, and halo statistics}

Here we introduce a few concepts and analytic results from the theory of 
structure formation which underly the use of galaxy clusters as cosmological 
probes. These provide us with the vocabulary which pervades the literature on 
analytic and numerical models of galaxy cluster evolution. Material in this 
section has been derived from three primary sources: Padmanabhan (1993) 
\cite{pad93} for 
the spherical top-hat model for nonlinear collapse, Dodelson (2003) 
\cite{dodelson03} for 
Press-Schechter theory, and Bryan {\&} Norman (1998) \cite{BN98}
for virial scaling relations.

\subsection{Nonlinearity defined}

In the linear regime, both super- and sub-horizon scale perturbations grow 
as $t^{2/3}$ in the matter-dominated era. This means that after recombination, 
the linear power spectrum retains its shape while its amplitude grows as 
$t^{4/3}$ before the onset of cosmic acceleration. When $\Delta ^2(k)$ for a 
given k approaches unity linear theory no longer applies, and some other 
method must be used to determine the fluctuation's growth. In general, 
numerical simulations are required to model the nonlinear phase of growth 
because in the nonlinear regime, the modes do not grow independently. 
Mode-mode coupling modifies both the shape and amplitude of the power 
spectrum over the range of wavenumbers that have gone nonlinear. 

At any given time, there is a critical wavenumber which we shall call the 
nonlinear wavenumber k$_{nl}$ which determines which portion of the spectrum 
has evolved into the nonlinear regime. Modes with k$<$k$_{nl}$ are said to 
be linear, while those for which k$>$ k$_{nl}$ are nonlinear. 
Conventionally, one defines the nonlinear wavenumber such that $\Delta 
(k_{nl} ,z)=1.$ From this one can derive a nonlinear mass scale $M_{nl} 
(z)=\frac{4\pi }{3}\bar {\rho }(z)\left( {\frac{2\pi }{k_{nl} }} \right)^3$. 
A more useful and rigorous definition of the nonlinear mass scale comes from 
evaluating the amplitude of mass fluctuations within spheres or radius R at 
epoch z. The enclosed mass is $M=\frac{4\pi }{3}\bar {\rho }(z)R^3.$ The mean 
square mass fluctuations (variance) is
\begin{equation}\label{eq17}
\left\langle {(\delta M/M)^2} \right\rangle \equiv \sigma ^2(M)=\int 
{d^3kW_T^2 (kR)P(k,z),} 
\end{equation}
where W is the Fourier transform of the top-hat window function
\begin{equation}\label{eq18}
\begin{array}{l}
 \mbox{W(}{\rm {\bf x}}\mbox{)}=\left\{ {{\begin{array}{*{20}c}
 {3/4\pi R^3,\mbox{ }\left| {\rm {\bf x}} \right|<R} \hfill \\
 {0,\mbox{ }\left| {\rm {\bf x}} \right|\ge R} \hfill \\
\end{array} }} \right. \\ 
 \to W_T (kR)=3\left[ {\sin (kR)/kR-\cos (kR)} \right]/(kR)^2. \\ 
 \end{array}
\end{equation}
If we approximate P(k) locally with a power-law $P(k,z)=D^2(z)k^m$, where D 
is the linear growth factor, then $\sigma ^2(M)\propto D^2R^{-(3+m)}\propto 
D^2M^{-(3+m)/3}.$ From this we see that the RMS fluctuations are a 
decreasing function of M. At very small mass scales, m$\rightarrow -3$, and 
the fluctuations asymptote to a constant value. We now define the nonlinear 
mass scale by setting $\sigma $(M$_{nl})$=1. We get that (\cite{white94})
\begin{equation}\label{eq19}
M_{nl} (z)\propto D(z)^{6/(3+m)}\mbox{ (}\propto 
\mbox{(1}+\mbox{z)}^{\mbox{-6/(3}+\mbox{m)}}\mbox{ for EdS).}
\end{equation}
For $m > -3$, the smallest mass scales become nonlinear first. This is the 
origin of hierarchical (``bottom-up'') structure formation. 

\subsection{Spherical Top-Hat Model}

\begin{figure}[htbp]
\centerline{\includegraphics[width=3in,height=2in]{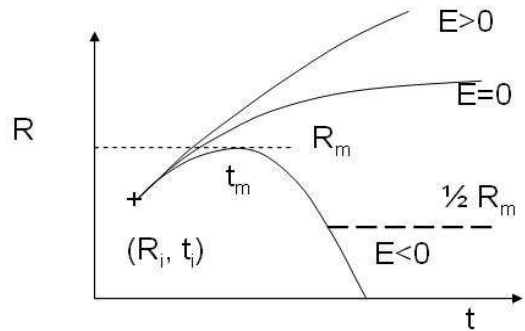}}
\caption{Evolution of a top-hat perturbation in an EdS universe. Depending 
on the E, the first integral of motion, the fluctuation collapses (E$<$0), 
continues to expand (E$>$0), or asymptotically reaches it maximum radius 
(E=0). Virialization occurs when the fluctuation has collapsed to half its 
turnaround radius.}
\label{fig5}
\end{figure}

We now ask what happens when a spherical volume of mass M and radius R 
exceeds the nonlinear mass scale. The simplest analytic model of the 
nonlinear evolution of a discrete perturbation is called the spherical 
top-hat model. In it, one imagines as spherical perturbation of radius $R$
and some constant overdensity $\bar {\delta }=3M/4\pi R^3$ in an Einstein-de Sitter 
(EdS) universe. By Birkhoff's theorem the equation of motion for R is 
\begin{equation}\label{eq20}
\frac{d^2R}{dt^2}=-\frac{GM}{R^2}=-\frac{4\pi G}{3}\bar {\rho }(1+\bar 
{\delta })R
\end{equation}
whereas the background universe expands according to Eq. \ref{eq6}
\begin{equation}\label{eq21}
\frac{d^2a}{dt^2}=-\frac{4\pi G}{3}\bar {\rho }a.
\end{equation}

Comparing these two equations, we see that the perturbation evolves like a 
universe of a different mean density, but with the same initial expansion 
rate. Integrating Eq. \ref{eq20} once with respect to time gives us the first 
integral of motion:
\begin{equation}\label{eq22}
\frac{1}{2}\left( {\frac{dR}{dt}} \right)^2-\frac{GM}{R}=E,
\end{equation}
where E is the total energy of the perturbation. If E$<$0, the perturbation 
is bound, and obeys
\begin{equation}\label{eq23}
\frac{R}{R_m}=\frac{(1-cos \theta)}{2}, ~~~\frac{t}{t_m}=\frac{(\theta-sin\theta)}{\pi}
\end{equation}
where $R_m$ and $t_m$ are the radius and time of ``turnaround''. At turnaround 
(as $\theta \rightarrow \pi$), the fluctuation reaches its maximum proper 
radius (see Fig. \ref{fig5}). As 
$t\rightarrow 2t_m, R\rightarrow 0$, and we say the fluctuation has collapsed.

A detailed analysis of the evolution of the top-hat perturbation is given in 
Padmanabhan (1993, Ch. 8) for general $\Omega_m$.
Here we merely quote results for an EdS universe.
The mean \textit{linear} overdensity at turnaround; i.e., the value one would predict from the linear growth formula $\delta \sim t^{2/3}$, is 1.063. The actual overdensity at 
turnaround using the nonlinear model is 4.6. This illustrates that nonlinear 
effects set in well before the amplitude of a linear fluctuation reaches 
unity. As R$\rightarrow $0, the nonlinear overdensity becomes infinite. 
However, the linear overdensity at $t=2t_m$ is only 1.686. As the fluctuation 
collapses, other physical processes (pressure, shocks, violent relation) 
become important which 
establish a gravitationally bound object in virial equilibrium before 
infinite density is reached. Within the framework of the spherical top-hat 
model, we say virialization has occurred when the kinetic and gravitational 
energies satisfy virial equilibrium: $\left| U \right|=2K.$ It is easy to 
show from conservation of energy that this occurs when $R=R_m/2$; in other 
words, when the fluctuation has collapsed to half its turnaround radius. The 
nonlinear overdensity at virialization $\Delta _c$
is not infinite since the radius is finite. 
For an EdS universe, $\Delta _c =18\pi ^2\approx 180$. Fitting
formulae for non-EdS models are provided in the next section.

\subsection{Virial Scaling Relations}
The spherical top-hat model can be scaled to perturbations of arbitrary 
mass. Using virial equilibrium arguments, we can predict various physical 
properties of the virialized object. The ones that interest us most are 
those that relate to the observable properties of gas in galaxy clusters, 
such as temperature, X-ray luminosity, and SZ intensity change. Kaiser \cite{kaiser86}
first derived virial scaling relations for clusters in an EdS universe. Here we 
generalize the derivation to non-EdS models of interest. In order to compute 
these scaling laws, we must assume some model for the distribution of matter 
as a function of radius within the virialized object. A top-hat distribution 
with a density $\rho =\Delta _c \bar {\rho }(z)$ is not useful because it is 
not in mechanical equilibrium. More appropriate is the isothermal, 
self-gravitating, equilibrium sphere for the collisionless matter, whose 
density profile is related to the one-dimensional velocity dispersion 
\cite{bt87}
\begin{equation}\label{eq24}
\rho (r)=\frac{\sigma ^2}{2\pi Gr^2}.
\end{equation}
If we define the virial radius r$_{vir}$ to be the radius of a spherical volume 
within which the mean density is $\Delta _{c}$ times the critical density 
at that redshift ($M=4\pi r_{vir}^3 \rho _{crit} \Delta _c /3)$, then there 
is a relation between the virial mass M and $\sigma $:
\begin{equation}
\label{eq25}
\sigma =M^{1/3}[H^2(z)\Delta _c G^2/16]^{1/6}\approx 476f_\sigma \left( 
{\frac{M}{10^{15}M_\odot }} \right)^{1/3}(h^2\Delta _c E^2)^{1/6}\mbox{ km 
s}^{\mbox{-1}}.
\end{equation}
Here we have introduced a normalization factor $f_{\sigma}$ which will be used to 
match the normailization from simulations. The redshift dependent Hubble 
parameter can be written as $H(z)=100hE(z)\mbox{ km s}^{-1}$ with the 
function $E^2(z)=\Omega _m (1+z)^3+\Omega _k (1+z)^2+\Omega _\Lambda $, 
where the $\Omega$'s have been previously defined. 

The value of $\Delta_c$ is taken from the spherical top-hat model, and is 18$\pi 
^{2}$ for the critical EdS model, but has a dependence on cosmology 
through the parameter $\Omega (z)=\Omega _m (1+z)^3/E^2(z).$ Bryan and Norman 
(1998) provided fitting formulae for $\Delta_c$ for the critical  for both open universe models and flat, lambda-dominated models
\begin{equation}\label{eq26}
\Delta _c =18\pi ^2+82x-39x^2\mbox{ for }\Omega _k =0,\mbox{ }\Delta _c 
=18\pi ^2+60x-32x^2\mbox{ for }\Omega _\Lambda =0
\end{equation}
where x=$\Omega $(z)-1. 

If the distribution of the baryonic gas is also isothermal, we can define a 
ratio of the ``temperature'' of the collisionless material ($T_\sigma =\mu 
m_p \sigma ^2/k)$ to the gas temperature:
\begin{equation}
\label{eq27}
\beta =\frac{\mu m_p \sigma ^2}{kT}
\end{equation}
Given equations (\ref{eq26}) and (\ref{eq27}), the relation between temperature and mass is then
\begin{equation}
\label{eq28}
kT=\frac{GM^{2/3}\mu m_p }{2\beta }\left[ {\frac{H^2(z)\Delta _c }{2G}} 
\right]^{1/3}\approx 1.39f_T \left( {\frac{M}{10^{15}M_\odot }} 
\right)^{2/3}(h^2\Delta _c E^2)^{1/3}\mbox{ keV,}
\end{equation}
where in the last expression we have added the normalization factor f$_{T}$ 
and set $\beta $=1.

The scaling behavior for the object's X-ray luminosity is easily computed by 
assuming bolometric bremsstrahlung emission and ignoring the temperature 
dependence of the Gaunt factor: $L_{bol} \propto 
\int {\rho ^2} T^{1/2}dV\propto M_b \rho T^{1/2}.$ where M$_{b}$ is the 
baryonic mass of the cluster. This is infinite for an isothermal density 
distribution, since $\rho $ is singular. Observationally and 
computationally, it is found that the baryon distribution rolls over to a 
constant density core at small radius. A procedure is described in Bryan and 
Norman (1998) which yields a finite luminosity:
\begin{equation}
\label{eq29}
L_{bol} =1.3\times 10^{45}\left( {\frac{M}{10^{15}M_\odot }} 
\right)^{4/3}(h^2\Delta _c E^2)^{7/6}\mbox{ }\left( {\frac{\Omega _b 
}{\Omega _m }} \right)^2\mbox{ erg s}^{-1}.
\end{equation}
Eliminating M in favor of T in Eq. \ref{eq29} we get
\begin{equation}
\label{eq30}
L_{bol} =6.8\times 10^{44}\left( {\frac{kT/f_T }{1.0\mbox{ keV}}} 
\right)^2(h^2\Delta _c E^2)^{1/2}\mbox{ }\left( {\frac{\Omega _b }{\Omega _m 
}} \right)^2\mbox{ erg s}^{-1}.
\end{equation}
The scaling of the SZ ``luminosity'' is likewise easily computed. If we 
define L$_{SZ}$ as the integrated SZ intensity change: $L_{SZ} =\int {dA\int {n_e 
\sigma _T } } \left( {\frac{kT}{m_e c^2}} \right)dl\propto M_b T$, then
\begin{equation}\label{eq30a}
L_{SZ} =\frac{GM^{5/3}\sigma _T }{2\beta m_e c^2}\left[ {\frac{H^2(z)\Delta 
_c }{2G}} \right]^{1/3}\left( {\frac{\Omega _b }{\Omega _m }} \right).
\end{equation}
We note that cosmology enters these relations only with the combination of 
parameters $h^2\Delta _c E^2$, which comes from the relation between the 
cluster's mass and the mean density of the universe at redshift z. The 
redshift variation comes mostly from E(z), which is equal to (1+z)$^{3/2}$ 
for an EdS universe. 

\subsection{Statistics of hierarchical clustering: Press-Schechter theory}
Now that we have a simple model for the nonlinear evolution of a spherical 
density fluctuation and its observable properties as a function of its 
virial mass, we would like to estimate the number of virialized objects of 
mass M as a function of redshift given the matter power spectrum. This is 
the key to using surveys of galaxy clusters as cosmological probes. While 
large scale numerical simulations can and have been used for this purpose 
(see below), we review a powerful analytic approach by Press and Schechter 
\cite{ps74} which turns out to be remarkably close to numerical results. The basic 
idea is to imagine smoothing the cosmological density field at any epoch z 
on a scale R such that the mass scale of virialized objects of interest 
satisfies $M=\frac{4\pi }{3}\bar {\rho }(z)R^3.$ Because the density field 
(both smoothed and unsmoothed) is a Gaussian random field, the probability 
that the mean overdensity in spheres of radius R exceeds a critical 
overdensity $\delta _{c}$ is
\begin{equation}\label{eq31}
p(R,z)=\frac{2}{\sqrt {2\pi } \sigma (R,z)}\int\limits_{\delta _c }^\infty 
{d\delta } \exp \left( {-\frac{\delta ^2}{2\sigma ^2(R,z)}} \right)
\end{equation}
where $\sigma(R,z)$ is the RMS density 
variation in spheres of radius R as discussed above. 
Press and Schechter suggested that this probability be identified with the 
fraction of particles which are part of a nonlinear lump with mass exceeding 
M if we take $\delta _c =1.686,$ the linear overdensity at virialization. 
This assumption has been tested against numerical simulations and found to 
be quite good \cite{wef93}. The fraction of the 
volume collapsed into objects with mass between $M$ and $M+dM$ is given by 
$(dp/dM)dM$. Multiply this by the average number density of such objects 
$\rho _m /M$ to get the number density of collapsed objects between 
$M$ and $M+dM$:
\begin{equation}\label{eq32}
dn(M,z)=-\frac{\bar {\rho }}{M}\frac{dp(M(R),z)}{dM}dM.
\end{equation}
The minus sign appears here because p is a decreasing function of M. 
Carrying out the derivative using the fact that $dM/dR=3M/R,$ we get
\begin{equation}\label{eq33}
\frac{dn(M,z)}{dM}=\sqrt {\frac{2}{\pi }} \frac{\bar {\rho }\delta _c 
}{3M^2\sigma }e^{-\delta _c^2 /2\sigma ^2}\left[ {-\frac{d\ln \sigma }{d\ln 
R}} \right].
\end{equation}
The term is square brackets is related to the logarithmic slope of the power 
spectrum, which on the mass scale of galaxy clusters is close to unity. Eq. 
\ref{eq33} is called the \textit{halo mass function}, and it has the form of a 
power law multiplied by an exponential. To make this more explicit, approximate the power spectrum on scales of interest as a power law as we have done above. Substituting the 
scaling relations for $\sigma $ in Eq. \ref{eq33} one gets the result \cite{white94}
\begin{equation}\label{eq34}
\frac{dn}{dM}=\left( {\frac{2}{\pi }} \right)^{1/2}\frac{\bar {\rho 
}}{M^2}\left( {1+\frac{m}{3}} \right)\left[ {\frac{M}{M_{nl} (z)}} 
\right]^{\frac{m-3}{6}}\exp \left[ {-\left( {\frac{M}{M_{nl} (z)}} 
\right)^{\frac{3+m}{3}}/2} \right].
\end{equation}
Here, $M_{nl} (z)$ is the nonlinear mass scale. To be more consistent with 
the spherical top-hat model, it satisfies the relation $\sigma (M_{nl} 
,z)=\delta _c $; i.e., those fluctuations in the smoothed density field that 
have reached the linear overdensity for which the spherical top-hat model 
predicts virialization.

\subsection{Application to galaxy clusters}

\begin{figure}[htbp]
\includegraphics[width=5in,height=3.33in]{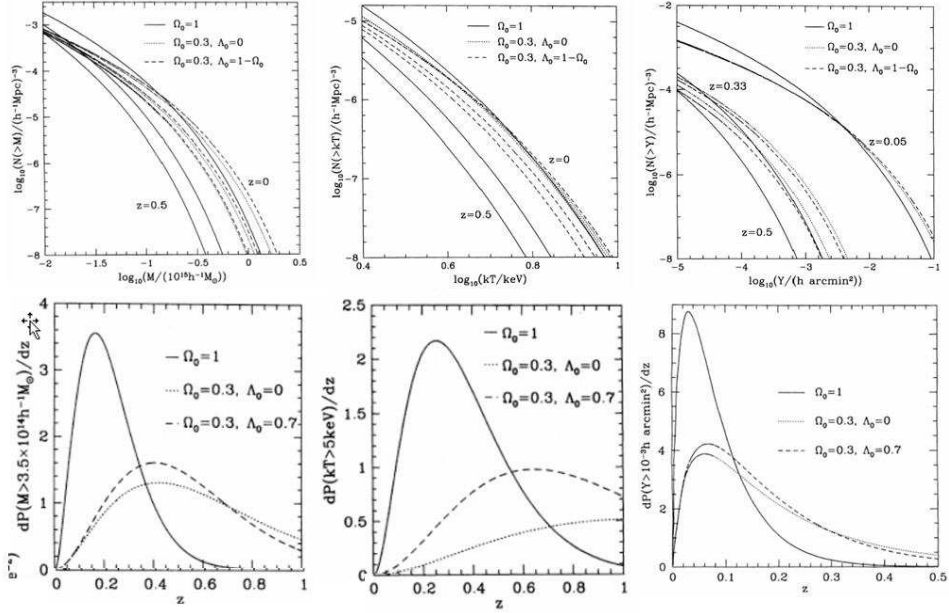}
\caption{Top left to bottom right: a) Integrated cluster mass function for 
three cosmologies and two redshifts; b) like a), but for integrated 
temperature function; c) like a) but for integrated SZ cross section; d) 
redshift distribution of the integrated probability to find a cluster 
exceeding $M=3.5 \times 10^{14} h^{-1} M_{\odot}$; e) redshift distribution of the 
integrated probability to find a cluster exceeding kT=5 keV; f) redshift 
distribution of the integrated probability to find a cluster exceeding 
Y=$10^{-3}$ h arcmin$^{2}$. From \cite{ecf96}.}
\label{fig6}
\end{figure}

Galaxy clusters correspond to rare ($\sim $3$\sigma )$ peaks in the density 
field. Combining the halo mass function 
as prediced by the PS formalism with the scaling laws derived above, we can 
predict the evolution of the statistical properties of X-ray and SZ clusters 
of galaxies. Here we show a few results taken from Eke, Cole {\&} Frenk 
(1996) \cite{ecf96}. 
Fig. 6a shows the evolution of the integrated mass function $n(>M)$ 
for several cosmologies and redshifts. One can see the power-law behavior at 
lower mass and the exponential cutoff at higher M. One sees strong redshift 
evolution of the number of massive clusters in the EdS model, but slower 
evolution on the open and lambda models. This is because of the saturated 
growth of structure in low density models. This makes 
number counts of massive clusters a sensitive test of the linear growth
factor D(z), which depends on $\Omega_m$ and $\Omega_{\Lambda}$. 
Convolving the cluster population with the 
scaling relations for T(M) and Y(M), one gets distribution functions for 
n($>$T) and n($>$Y). Here $Y=L_{SZ}/d_A^2$ is the effective SZE cross section of a cluster, where $d_A$ is its angular diameter distance.
These are shown in Figs. 6b and 6c. Another way to 
present the data is to convolve the mass function with the differential 
volume element as a function of redshift for the three models. Figs. 6d-f 
plot the redshift probability of detecting a cluster with M, T, and Y 
exceeding the fiducial values given in the figure caption. As one can see, 
the profiles are sharply peaked at low redshift for the EdS model, but 
substantially broader and peaking at higher redshift for the low density 
universe models. There is, however, rather little difference between the 
open and lambda-dominated models as far as the probability distributions for 
M and Y. Things are somewhat better for T, implying that some combination of 
X-ray and SZE measurements will be needed for precision 
cosmological parameter determinations.

\section{Numerical simulations of gas in galaxy clusters}

The central task is for a given cosmological model, calculate the formation 
and evolution of a population of clusters from which synthetic X-ray and SZ 
catalogs can be derived. These can be used to calibrate simpler analytic 
models, as well as to build synthetic surveys (mock catalogs) which can be 
used to assess instrumental effects and survey biases. One would like to 
directly simulate $n(M,z), n(L_x,z), n(T,z), n(Y,z)$ from the governing 
equations for collisionless and collisional matter in an expanding universe. 
Clearly, the quality of these statistical predictions relies on the ability to 
adequately resolve the internal structure and thermodynamical evolution of 
the ICM. 

In Norman (2003) \cite{norman03}
I provided a historical review of the progress that has 
been made in simulating the evolution of gas in galaxy clusters motivated by 
X-ray observations. Since X-ray emission and the SZE are both consequences 
of hot plasma bound in the cluster's gravitational potential well, the 
requirements to faithfully simulate X-ray clusters and SZ clusters are 
essentially the same. Numerical progress can be characterized as a quest for 
higher resolution and essential baryonic physics. In this section I describe 
the technical challenges involved and the numerical methods that have been 
developed to overcome them. I then discuss the effects of assumed baryonic 
physics on ICM structure. Our point of reference is the non-radiative (so-called 
adiabatic) case, which has been the subject of an extensive code comparison 
\cite{Frenk99}. I review the properties of adiabatic X-ray clusters, 
and show that they fail to reproduce observed cluster scaling laws. I then 
show results of numerical hydrodynamic simulations incorporating radiative 
cooling, star formation, and galaxy feedback and their associated scaling 
properties.

\subsection{Dynamic range considerations}

\begin{figure}[htbp]
\includegraphics[width=3in,height=1.7in]{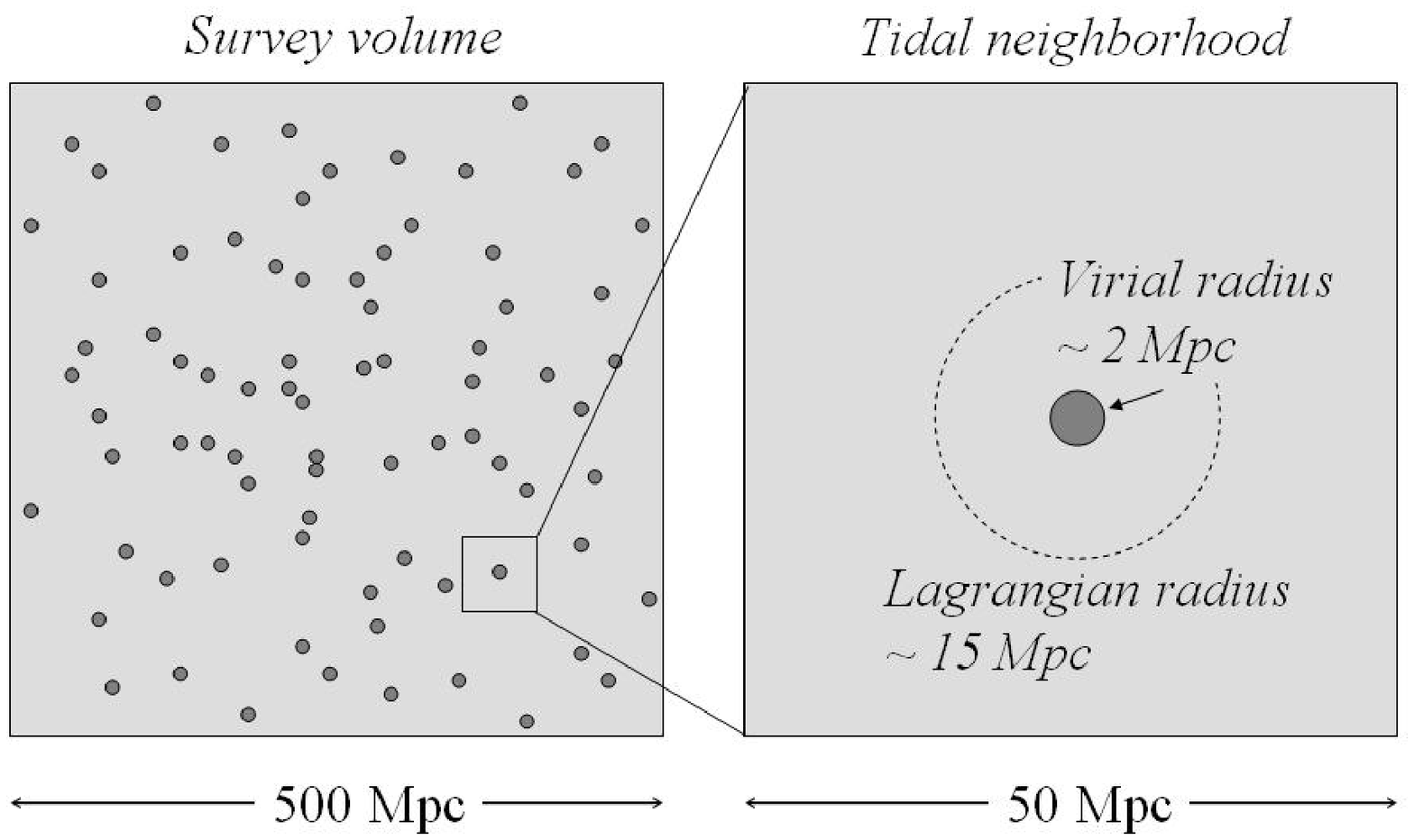}
\includegraphics[width=2.3in,height=1.5in]{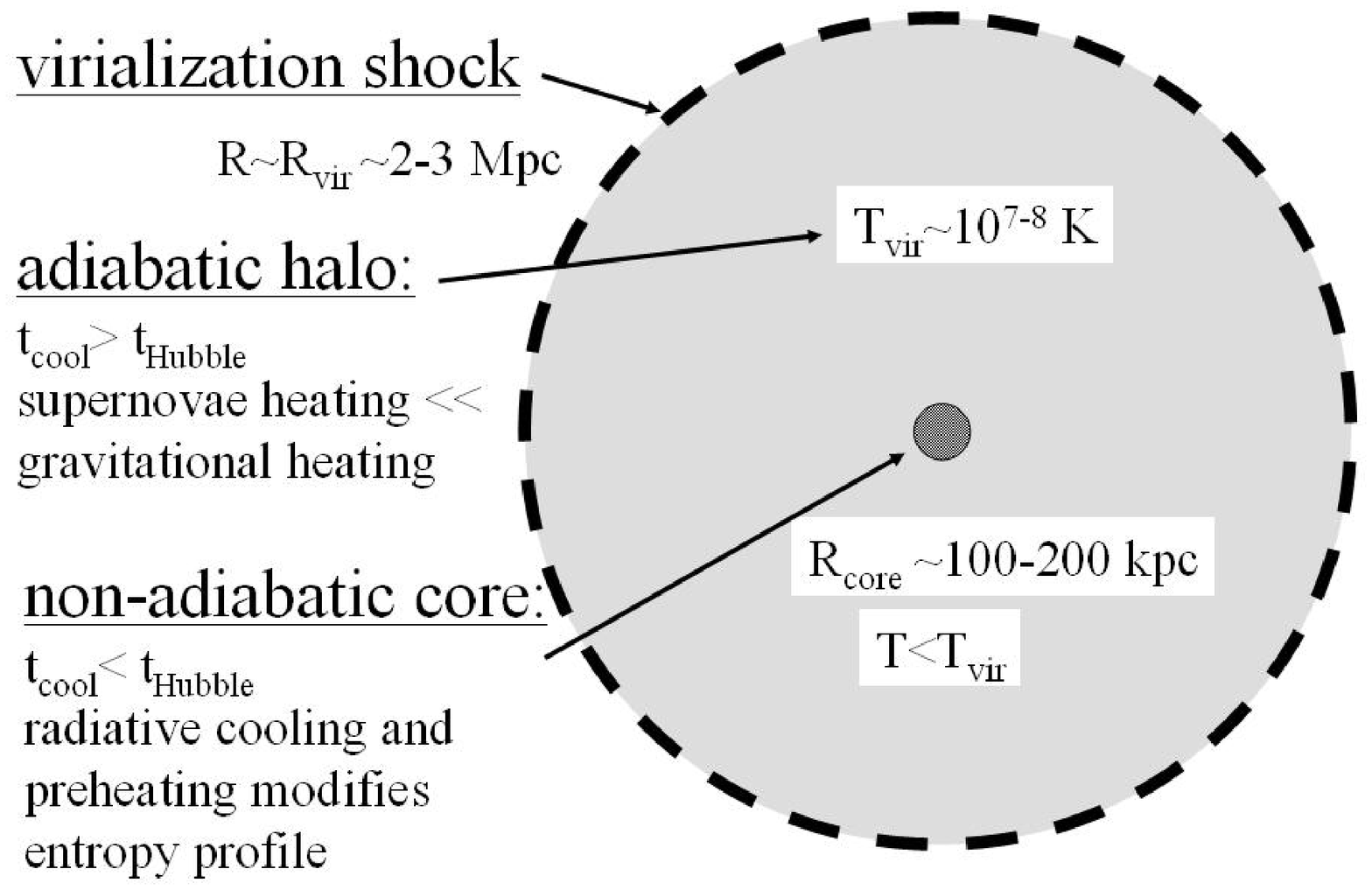}
\caption{Left: A range of length scales of $\sim $250 separates the size of a reasonable survey volume and the virial radius of a rich cluster. 
Right: Simplified structure of the ICM in a massive cluster. A range of length scales of $\sim $20-30 separates the virial radius and the core radius. }
\label{fig7}
\end{figure}

Figure 7 illustrates the dynamic range difficulties encountered with 
simulating a statistical ensemble of galaxy clusters, while at the same time 
resolving their internal structure. Massive clusters are rare at any 
redshift, yet these are the ones most that are most sensitive to cosmology. 
From the cluster mass function (Fig. 6a), in order to get adequate 
statistics, one deduces that one must simulate a survey volume many hundreds 
of megaparsecs on a side (Fig. 7a). A massive cluster has a virial radius of 
$\sim $2 Mpc. It forms via the collapse of material within a comoving 
Lagrangian volume of $\sim $15 Mpc. However, tidal effects from a larger 
region (50-100 Mpc) are important on the dynamics of cluster formation. The 
internal structure of cluster's ICM is shown in Fig. 7b. While clusters are 
not spherical, two important radii are generally used to characterize them: 
the virial radius, which is the approximate location of the virialization 
shock wave that thermalizes infalling gas to 10-100 million K, and the core 
radius, within which the baryon densities plateau and the highest X-ray 
emissions and SZ intensity changes are measured. A typical radius is $\sim $200 
kpc. Within the core, radiative cooling and possibly other physical 
processes are important. Outside the core, cooling times are longer than the 
Hubble time, and the ICM gas is effectively adiabatic. If we wanted to achieve 
a spatial resolution of 1/10 of a core radius everywhere within the survey 
volume, we would need a spatial dynamic range of D=500 Mpc/20 kpc = 25,000. 
The mass dynamic range is more severe. If we want 1 million dark matter 
particles within the virial radius of a $10^{15} M_{\odot}$ cluster, then we would 
need $N_{particle} =M_{box} /M_{particle} =\Omega _m \rho _{crit} 
L^3/10^9\approx 10^{11}$ if they were uniformly distributed in the survey 
volume. 

Two solutions to spatial dynamic range problem have been developed: tree 
codes for gridless N-body methods \cite{KWH96,syw01} 
and adaptive mesh refinement (AMR) for Eulerian particle-mesh/hydrodynamic methods \cite{bn97,Kravtsov97,Teyssier02,OShea04}. 
Both methods increase the spatial 
resolution automatically in collapsing regions as described below. The 
solution to the mass dynamic range problem is the use of multi-mass initial 
conditions in which a hierarchy of particle 
masses is used, with many low mass particles concentrated in the region of 
interest. This approach has most recently used by Springel et al. (2000)
\cite{springel00}, 
who simulated the formation of a galaxy cluster dark matter halo with 
$N=6.9\times 10^6$ dark matter particles, resolving the dark matter halos 
down to the mass scale of the Fornax dwarf spheroidal galaxy. The spatial 
dynamic range achieved in this simulation was $R=2\times 10^5$. Such dynamic 
ranges have not yet been achieved in galaxy cluster simulations with gas.

\subsection{Simulating cluster formation}

Simulations of cosmological structure formation are done in a cubic domain 
which is comoving with the expanding universe. Matter density and velocity 
fluctuations are initialized at the starting redshift chosen such that all 
modes in the volume are still in the linear regime. 
Once initialized, these fluctuations are then evolved to 
z=0 by solving the equations for collisionless N-body dynamics for cold dark 
matter, and the equations of ideal gas dynamics for the baryons in an 
expanding universe. Making the transformation from proper to comoving 
coordinates $\vec {r}=a(t)\vec {x}$, Newton's laws for the collsionless dark 
matter particles become
\begin{equation}
\label{eq35}
\frac{d\vec {x}_{dm} }{dt}=\vec {\upsilon }_{dm} ,\mbox{ }\frac{d\vec 
{\upsilon }_{dm} }{dt}=-2\frac{\dot {a}}{a}\vec {\upsilon }_{dm} 
-\frac{1}{a^2}\nabla _x \phi 
\end{equation}
where $x$ and $v$ are the particle's comoving position and peculiar velocity, 
respectively, and $\phi$ is the comoving gravitational potential that includes 
baryonic and dark matter contributions. The hydrodynamical equations for 
mass, momentum, and energy conservation in an expanding universe in comoving 
coordinates are (\cite{Anninos97})
\begin{equation}
\label{eq36}
\begin{array}{l}
 \frac{\partial \rho _b }{\partial t}+\nabla \cdot (\rho _b \vec {\upsilon 
}_b )+3\frac{\dot {a}}{a}\rho _b =0, \\ 
 \frac{\partial (\rho _b \upsilon _{b,i} )}{\partial t}+\nabla \cdot [(\rho 
_b \upsilon _{b,i} )\vec {\upsilon }_b +5\frac{\dot {a}}{a}\rho _b \upsilon 
_{b,i} =-\frac{1}{a^2}\frac{\partial p}{\partial x_i }-\frac{\rho _b 
}{a^2}\frac{\partial \phi }{\partial x_i }, \\ 
 \frac{\partial e}{\partial t}+\nabla \cdot (e\vec {\upsilon }_b )+p\nabla 
\cdot \vec {\upsilon }_b +3\frac{\dot {a}}{a}e=\Gamma -\Lambda , \\ 
 \end{array}
\end{equation}
where $\rho_b, p$ and $e$, are the baryonic density, pressure and internal energy 
density defined in the proper reference frame, $\vec {\upsilon }_b $ is the 
comoving peculiar baryonic velocity, $a=1/(1+z)$ is the cosmological scale 
factor, and $\Gamma $ and $\Lambda $ are the microphysical heating and 
cooling rates. The baryonic and dark matter components are coupled through 
Poisson's equation for the gravitational potential
\begin{equation}
\label{eq37}
\nabla ^2\phi =4\pi Ga^2(\rho _b +\rho _{dm} -\bar {\rho }(z))
\end{equation}
where $\bar {\rho }(z)=3H_0 \Omega _m (0)/8\pi Ga^3$ is the proper 
background density of the universe.

The cosmological scale factor $a(t)$ is obtained by integrating the 
Friedmann equation (Eq. \ref{eq4}). To complete the specification of the problem we 
need the ideal gas equation of state $p=(\gamma -1)e$, and the gas heating 
and cooling rates. When simulating the ICM, the simplest approximation is to 
assume $\Gamma $ and $\Lambda =0$; i.e., no heating or cooling of 
the gas other than by adiabatic processes and shock heating. 
Such simulations are referred to as 
adiabatic (despite entropy-creating shock waves), and are a reasonable first approximation to real clusters because 
except in the cores of clusters, the radiative cooling time is longer than a 
Hubble time, and gravitational heating is much larger than sources of 
astrophysical heating. However, as discussed in the paper by Cavaliere in 
this volume, there is strong evidence that the gas in cores of clusters has 
evolved non-adiabatically. This is revealed by the entropy profiles observed 
in clusters \cite{Ponman99} which deviate substantially from adiabatic 
predictions. In the simulations presented below, we consider radiative 
cooling due to thermal bremsstrahlung, and mechanical heating due to galaxy 
feedback, details of which are described below.

\subsection{Numerical methods overview}

A great deal of literature exists on the gravitational clustering of CDM 
using N-body simulations. A variety of methods have been employed including 
the fast grid-based methods particle-mesh (PM), and 
particle-particle+particle-mesh (P$^{3}$M) \cite{Efstathiou81}, 
spatially adaptive methods such as adaptive P$^{3}$M \cite{Couchman91}, 
adaptive mesh refinement \cite{Kravtsov97}, tree codes 
\cite{BarnesHut86,WarrenSalmon94}, and hybrid methods such as TreePM 
\cite{Xu99}. Because of the large dynamic range required, 
spatially adaptive methods are favored, with Tree and TreePM methods the 
most widely used today. When gas dynamics is included, only certain 
combinations of hydrodynamics algorithms and collisionless N-body algorithms 
are ``natural''. Dynamic range considerations have led to two principal 
approaches: P$^{3}$MSPH and TreeSPH, which marries a P$^3$M or tree code for 
the dark matter with the Lagrangian smoothed-particle-hydrodynamics (SPH) 
method \cite{Evrard88,KWH96,syw01}, and adaptive mesh refinement (AMR), 
which marries PM with 
Eulerian finite-volume gas dynamics schemes on a spatially adaptive mesh 
\cite{bn97,OShea04,Teyssier02,Kravtsov03}.
Pioneering hydrodynamic simulations 
using non-adaptive Eulerian grids 
\cite{Kang94,Bryan94,BN98}
yielded some important insights about cluster formation and 
statistics, but generally have inadequate resolution to resolve their internal 
structure in large survey volumes. In the following we concentrate on our 
latest results using the AMR code \textit{Enzo} \cite{OShea04}. 
The reader is also 
referred to the paper by Borgani et al. \cite{Borgani04} which presents recent, 
high-resolution results from a large TreeSPH simulation.

\textit{Enzo} is a grid-based hybrid code (hydro + N-body) which uses the 
block-structured AMR algorithm of Berger {\&} Collela \cite{Berger89} to improve 
spatial resolution in regions of large gradients, such as in gravitationally 
collapsing objects. The method is attractive for cosmological applications 
because it: (\ref{eq1}) is spatially- and time-adaptive, (\ref{eq2}) uses accurate and 
well-tested grid-based methods for solving the hydrodynamics equations, and 
(\ref{eq3}) can be well optimized and parallelized. The central idea behind AMR is 
to solve the evolution equations on a grid, adding finer meshes in regions 
that require enhanced resolution. Mesh refinement can be continued to an 
arbitrary level, based on criteria involving any combination of overdensity 
(dark matter and/or baryon), Jeans length, cooling time, etc., enabling us 
to tailor the adaptivity to the problem of interest. The code solves the 
following physics models: collisionless dark matter and star particles, 
using the particle-mesh N-body technique \cite{Hockney88}; gravity, using FFTs on the 
root grid and multigrid relaxation on the subgrids; cosmic expansion; gas 
dynamics, using the piecewise parabolic method (PPM)\cite{Collela84}; 
multispecies 
nonequilibrium ionization and H$_{2}$ chemistry, using backward Euler time 
differencing \cite{Anninos97}; radiative heating and cooling, using subcycled forward 
Euler time differencing 
\cite{Anninos94}; and a parameterized star formation/ feedback 
recipe \cite{Cen92}. At the present time, magnetic fields and radiation transport 
are being installed. \textit{Enzo} is publicly available at
{\textit{http://cosmos.ucsd.edu/enzo}}.

\subsection{Structure of nonradiative clusters: the Santa Barbara test cluster}

In Frenk et al. \cite{Frenk99} 12 groups compared the results of a variety of 
hydrodynamic cosmological algorithms on a standard test problem. The test 
problem, called the Santa Barbara cluster, was to simulate the formation 
of a Coma-like cluster in a standard CDM cosmology ($\Omega_m=1$) 
assuming the gas is nonradiative. Groups 
were provided with uniform initial conditions and were asked to carry out a 
``best effort'' computation, and analyze their results at z=0.5 and z=0 for 
a set of specified outputs. These outputs included global integrated quantities, 
radial profiles, and column-integrated images. The simulations varied 
substantially in their spatial and mass resolution owing to algorithmic and 
hardware limitations. Nonetheless, the comparisons brought out which 
predicted quantities were robust, and which were not yet converged. In Fig. 
8 we show a few figures from Frenk et al. (1999) which highlight areas of 
agreement (top row) and disagreement (bottom row). 

\begin{figure}[htbp]
\includegraphics[width=5in,height=3.33in]{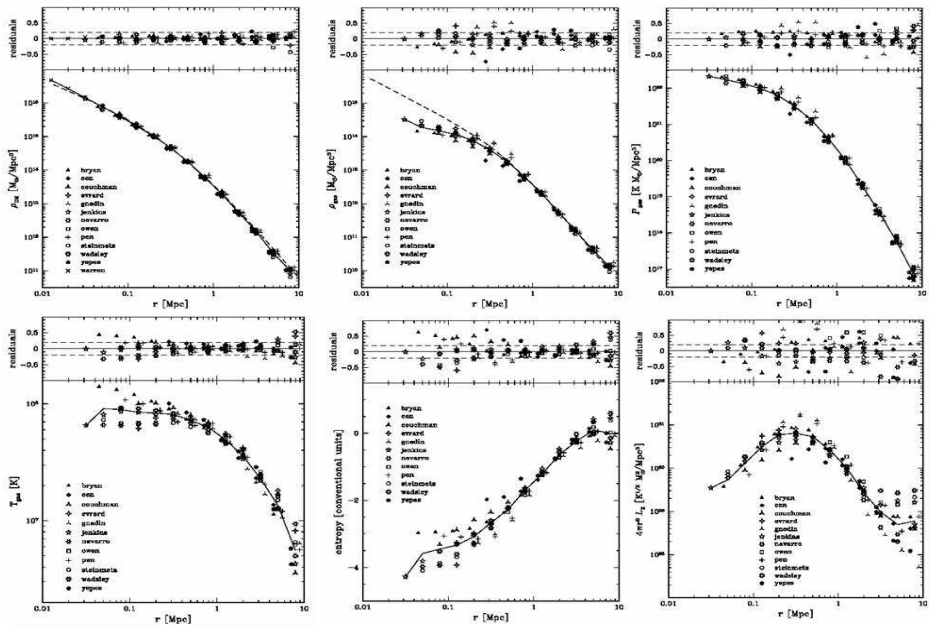}
\caption{The Santa Barbara test cluster. Top row, left to right: profiles
of dark matter density, gas density, and gas pressure. Bottom row, left
to right: profiles of gas temperature, gas entropy, and X-ray emissivity.
Different symbols correspond to different code results. From \cite{Frenk99}.}
\label{fig8}
\end{figure}

The top row shows profile of dark matter density, baryon density, and 
pressure for the different codes. All are in quite good agreement for the 
\textit{mechanical structure} of the cluster. The dark matter profile is well described by an NFW profile 
which has a central cusp \cite{NFW96}. The baryon 
density profiles show more dispersion, but all codes agree that the profile 
flattens at small radius, as observed. All codes agree extremely well on the 
gas pressure profile, which is not surprising, since mechanical equilibrium 
is easy to achieve for all methods even with limited resolution. This bodes 
well for the interpretation of SZE observations of clusters, since the 
Compton y parameter is proportional to the projected pressure distribution. 
In section 5 we show results from a statistical ensemble of clusters which 
bear this out. 

The bottom row shows the thermodynamic structure of the cluster, as well as 
the profile of X-ray emissivity. The temperature profiles show a lot of 
scatter within about one-third the virial radius (=2.7 Mpc). 
Systematically, the SPH codes produce nearly isothermal cores, while the 
grid codes produce temperature profiles which continue to rise as 
r$\rightarrow $0. The origin of this discrepancy has not been resolved, but 
improved SPH formulations come closer to reproducing the AMR results
\cite{Ascasibar03}. This discrepancy is reflected in the entropy 
profiles. Again, agreement is good in the outer two-thirds of the cluster, but the 
profiles show a lot of dispersion in the inner one third. Discounting the codes 
with inadequate resolution, one finds the SPH codes produce an entropy 
profile which continues to fall as r$\rightarrow $0, while the grid codes 
show an entropy core, which is more consistent with observations \cite{Ponman99}.
The dispersion in the density and temperature profiles are 
amplified in the X-ray emissivity profile, since $\varepsilon _x \propto 
\rho _b^2 T^{1/2}$. The different codes agree on the integrated X-ray luminosity of 
the cluster only to within a factor of 2. This is primarily because the 
density profile is quite sensitive to resolution in the core; any 
underestimate in the core density due to inadequate resolution is amplified 
by the density squared dependence of the emissivity. This suggests that 
quite high resolution is needed, as well as a good grasp on non-adiabatic 
processes operating in cluster cores, before simulations will be able to 
accurately predict X-ray luminosities. 

\subsection{A numerical sample of adiabatic clusters: Universal Temperature 
Profile}

Three questions one can ask about the Santa Barbara cluster results are: 1) 
is the cluster statistically representative, 2) do the results change 
substantially for a $\Lambda $CDM cosmology (the SB cluster assumed an EdS 
cosmology), and 3) what is the effect of additional baryonic physics on 
cluster structure? We address these questions here by summarizing results of 
\textit{Enzo} simulations of the ICM in a sample of clusters in a concordance $\Lambda$CDM model 
drawn from a survey volume 256h$^{-1}$ Mpc on a side. Multimass initial 
conditions and AMR are used to achieve high spatial and mass resolution 
within the clusters. More details can be found in \cite{Loken02,Motl04,Motl05,
Hallman05}.

\begin{figure}[htbp]
\includegraphics[width=2.5in,height=1.7in]{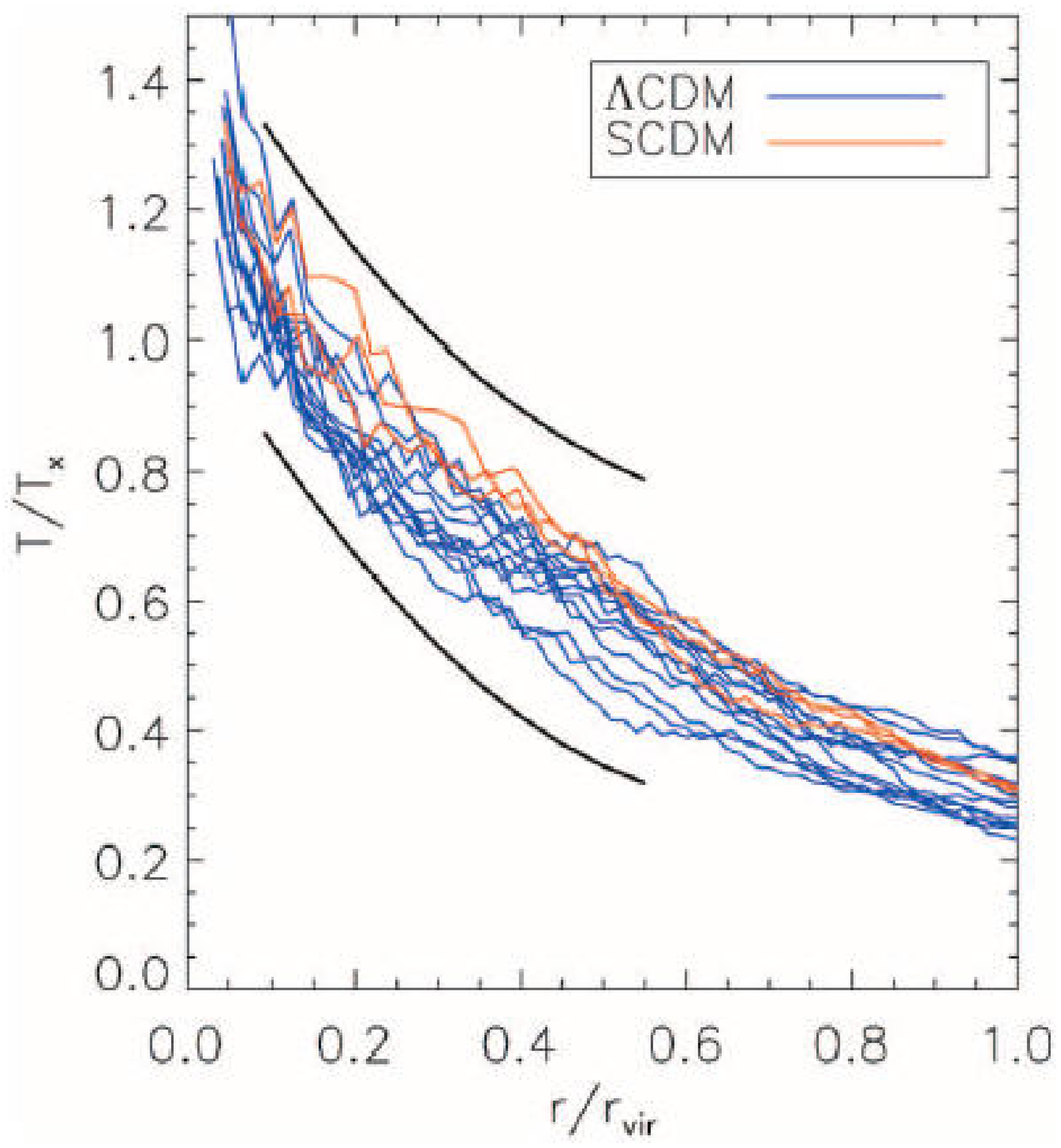}
\includegraphics[width=2.5in,height=1.7in]{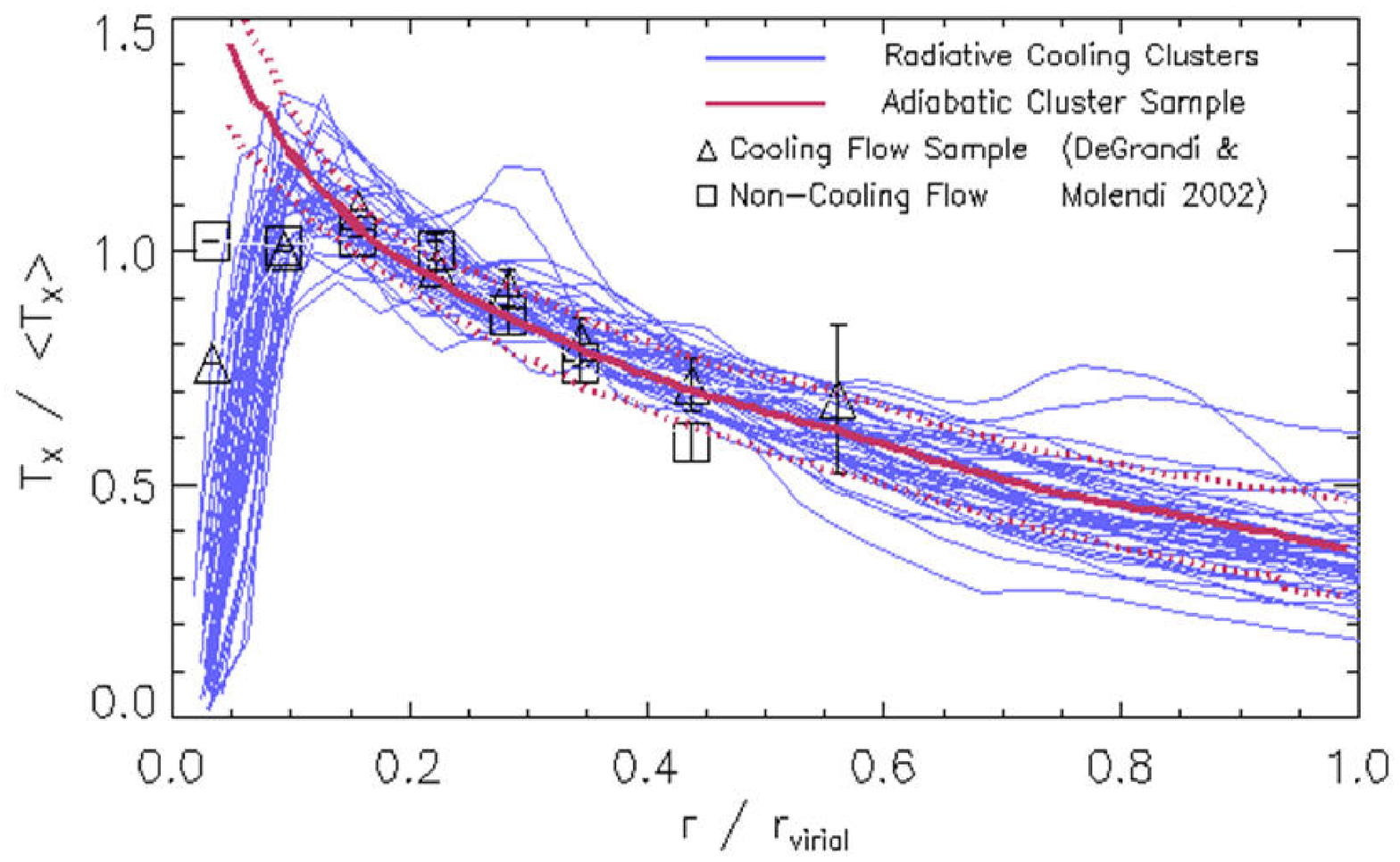}
\caption{Left: Temperature profiles from a sample of adiabatic cluster simulations (from Loken et al. 2002). Black curves bound the 1s confidence band from Markevitch et al. (1998). Right: Effect of radiative cooling on temperature profiles, compared with adiabatic sample average (red line) and observational data for cooling flow clusters (triangles) and non-cooling flow clusters (squares).}
\label{fig9}
\end{figure}

Fig. \ref{fig9} shows spherically averaged temperature profiles for 13(3) 
$\Lambda$CDM(SCDM) 
simulated clusters at z=0 analyzed by Loken et al. (2002)
 \cite{Loken02}. These were chosen 
from a total sample of 22(10) clusters because their 2D projected 
temperature maps were symmetric; the rejected non-symmetric clusters were in 
various states of merging. The smooth black curves bound the 1$\sigma $ 
confidence band from Markevitch et al. (1998)\cite{Markevitch98}
who analyzed temperature 
profiles from a sample of 17 symmetric X-ray clusters observed with ASCA. 
When temperature is normalized by the integrated emission-weighted 
temperature and the radius by the virial radius, both the observed data and 
the simulated data collapse to a narrow band, suggesting a universal 
temperature profile (UTP) outside the core region. 
The fit to the numerical data is $T\propto 
(1+r/\alpha )^{-\delta }$, with $\alpha \sim $r$_{vir}$/1.5 and $\delta 
\sim $1.6. The $\Lambda$CDM clusters and SCDM clusters exhibit the same profile, 
with a suggestion of a slightly higher normalization for clusters in the 
critically closed model. The fit is in good agreement with observations over 
the range 0.2$<$r/r$_{vir}<$0.5, but diverges at small radius where the 
effects of non-adiabatic processes appear to be at play \cite{deGrandi02}. 
The reality of the UTP was somewhat controversial when early 
results from Newton/XMM were showing large isothermal cores. However, the 
latest Chandra observations of 13 nearby, relaxed clusters have shown that 
the UTP provides an excellent description for temperature profiles outside 
$r\sim 0.15r_{vir}$ \cite{Vikhlinin04}. Subsequent numerical studies by 
Ascasibar et al. \cite{Ascasibar03} and Borgani et al. \cite{Borgani04}
using SPH have found 
agreement with the AMR results of Loken et al. The general agreement 
of numerical and observational results suggests that the declining 
temperature profile is a natural consequence of gravitational heating of the 
ICM during the process of cluster formation.

\subsection{Effect of additional physics}

\begin{figure}[htbp]
\includegraphics[width=3in,height=3.5in]{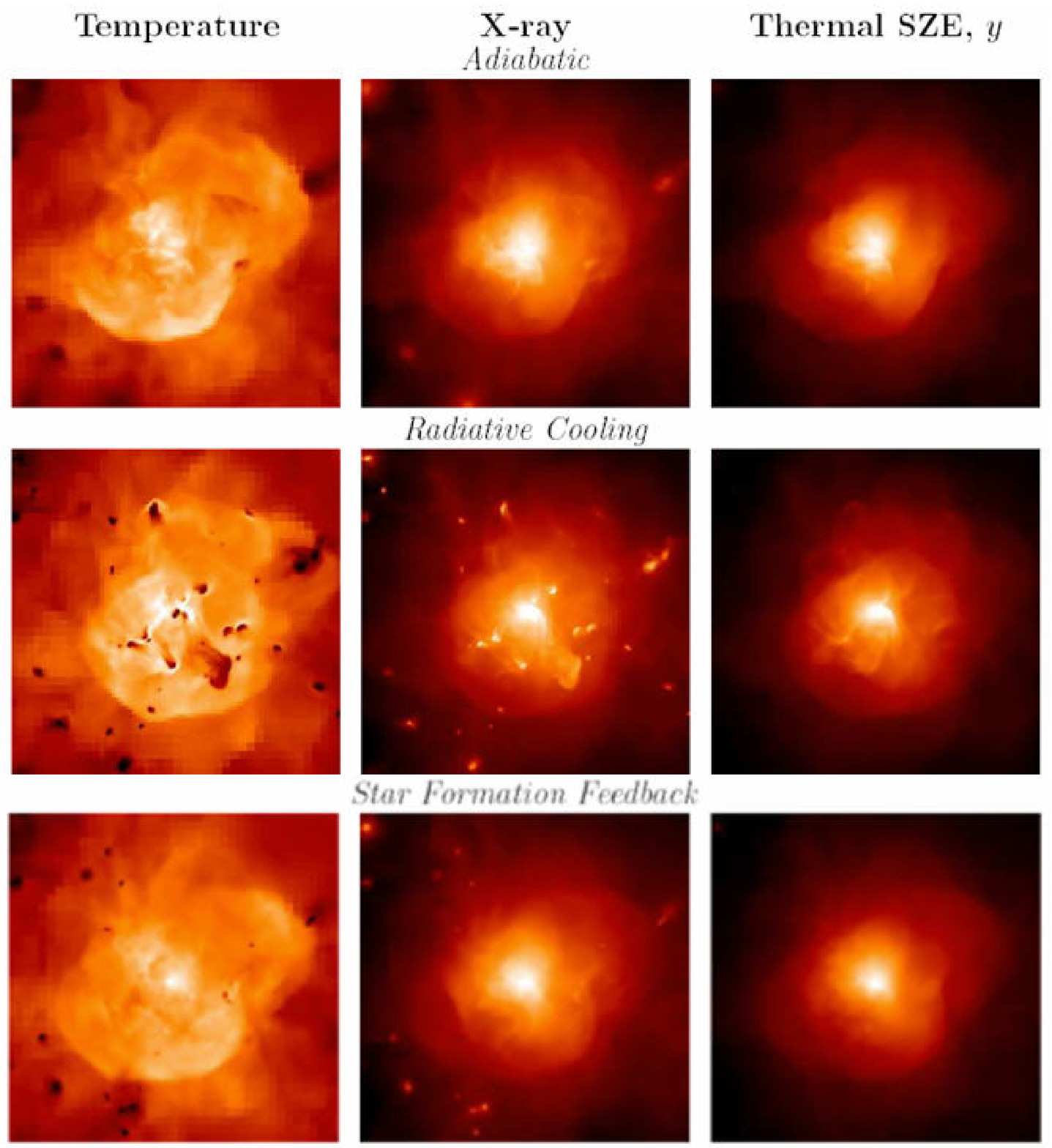}
\includegraphics[width=2.5in,height=2in]{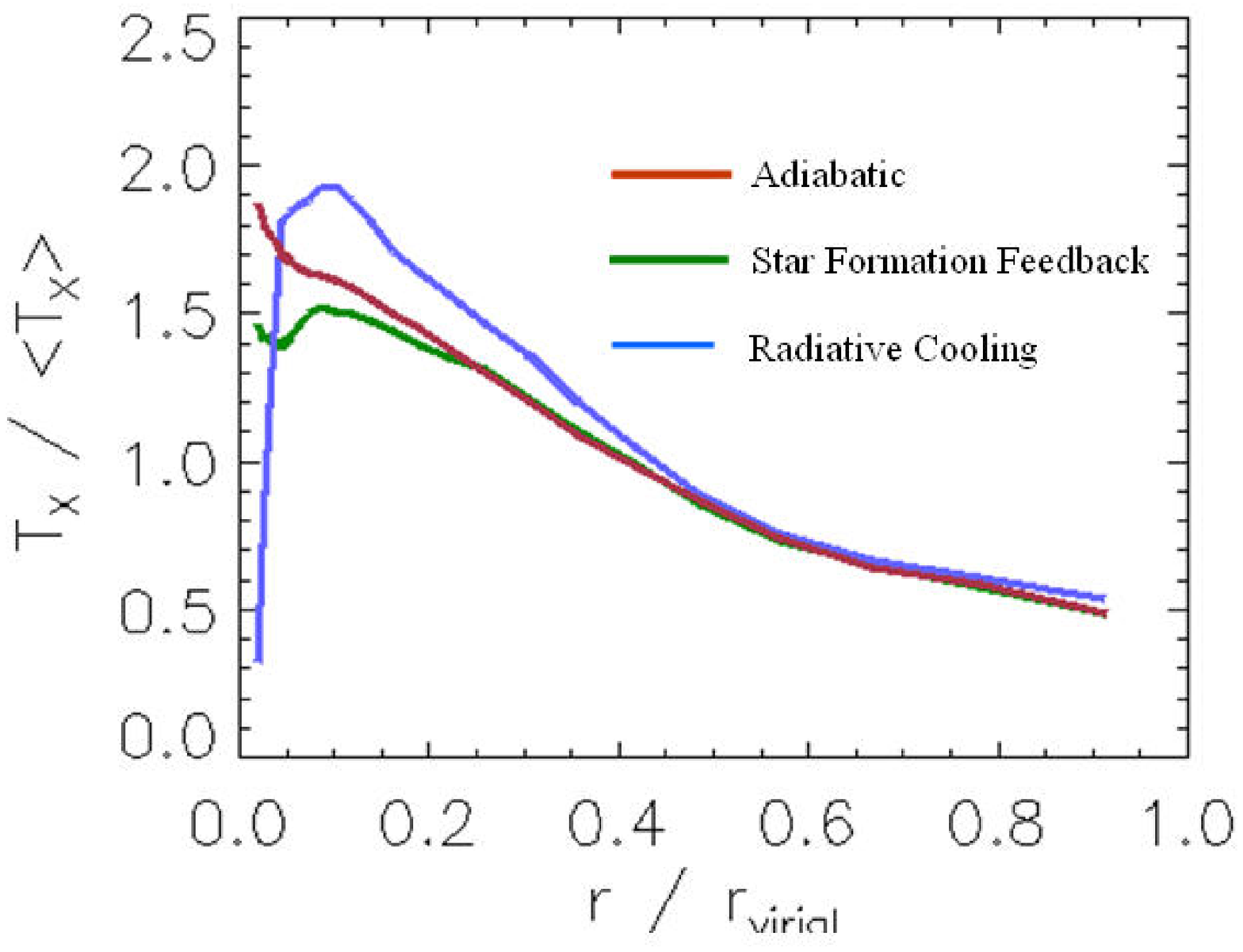}
\caption{Left: Columns show X-ray surface brightness, projected temperature, and Compton y-parameter for a $M=2\times 10^{15} M_{\odot}$ cluster assuming different baryonic physics. Field of view is 5 h$^{-1}$ Mpc. Right:
Corresponding spherically averaged radial temperature profiles.}
\label{fig10}
\end{figure}

Within r=0.15 r$_{vir}$, Vikhlinin et al. \cite{Vikhlinin04}
found large variation in 
temperature profiles, but in all cases the gas is cooler than the cluster 
mean. This suggests that radiative cooling is important in cluster cores, 
and possibly other effects as well. It has been long known that $\sim 
60$ percent of nearby, luminous X-ray clusters have central X-ray excesses, 
which has been interpreted as evidence for the presence of a cluster-wide 
cooling flows \cite{Fabian94}. More recently, Ponman et al. \cite{Ponman99}
have used X-ray 
observations to deduce the entropy profiles in galaxy groups and clusters. 
They find an entropy floor in the cores of clusters indicative of extra, 
non-gravitational heating, which they suggest is feedback from galaxy 
formation. It is easy to imagine cooling and heating both may be important 
to the thermodynamic evolution of ICM gas. 

To explore the effects of additional physics on the ICM, we recomputed the 
entire sample of clusters changing the assumed baryonic physics, keeping 
initial conditions the same. Three additional samples of about 100 clusters 
each were simulated: The ``radiative cooling'' sample assumes no additional 
heating, but gas is allowed to cool due to X-ray line and bremsstrahlung 
emission in a 0.3 solar metallicity plasma. The ``star formation'' sample 
uses the same cooling, but additionally cold gas is turned into 
collisionless star particles at a rate $\dot {\rho }_{SF} =\varepsilon _{sf} 
\frac{\rho _b }{\max (\tau _{cool} ,\tau _{dyn} )}$ , where $\varepsilon 
_{sf}$ is the star formation efficiency factor $\sim $0.1, and $\tau 
_{cool}$ and $\tau _{dyn}$ are the local cooling time and freefall time, 
respectively. This locks up cold baryons in a non-X-ray emitting component, 
which has been shown to have an important effect of the entropy profile of 
the remaining hot gas \cite{Bryan99,Voit00}. Finally, we have 
the ``star formation feedback'' sample, which is similar to the previous 
sample, except that newly formed stars return a fraction of their rest mass 
energy as thermal and mechanical energy. The source of this energy is high 
velocity winds and supernova energy from massive stars. In \textit{Enzo}, we implement 
this as thermal heating in every cell forming stars: $\Gamma _{sf} 
=\varepsilon _{SN} \dot {\rho }_{SF} c^2$. The feedback parameter depends on 
the assumed stellar IMF the explosion energy of individual supernovae. It is 
estimated to be in the range $10^{-6}\le \varepsilon _{SN} \le 10^{-5}$ \cite{Cen92}. We treat it as a free parameter.

Fig. \ref{fig10} shows synthetic maps of X-ray surface brightness, temperature, and 
Compton y-parameter for a $M=2\times 10^{15} M_{\odot}$ cluster at z=0 for the 
three cases indicated. The ``star formation'' case is omitted because the 
images are very similar to the ``star formation feedback'' case (see reference
\cite{Motl05}.) The adiabatic cluster shows that the X-ray emission is highly 
concentrated to the cluster core. The projected temperature distribution 
shows a lot of substructure, which is true for the adiabatic sample as a 
whole \cite{Loken02}. A complex virialization shock is toward the edge 
of the frame. The y-parameter is smooth, relatively symmetric, and centrally 
concentrated. The inclusion of radiative cooling has a strong effect on the 
temperature and X-ray maps, but relatively little effect on the SZE map. The 
significance of this is discussed in Section 5. In simulations with 
radiative cooling only, dense gas in merging subclusters cools to 10$^{4}$ K and 
is brought into the cluster core intact \cite{Motl04}. These cold lumps 
are visible as dark spots in the temperature map. They appear as X-ray 
bright features. The inclusion of star formation and energy feedback erases 
these cold lumps, producing maps in all three quantities that resemble 
slightly smoothed versions of the adiabatic maps. However, an analysis of 
the radial temperature profiles (Fig. \ref{fig10}) reveal important differences in 
the cluster core. The temperature continues to rise toward smaller radii in 
the adiabatic case, while it plummets to $\sim $10$^{4}$ K for the radiative 
cooling case. While the temperature profile looks qualitatively similar to 
observations of so-called cooling flow clusters, our central temperature is 
too low and the X-ray brightness too high. The star 
formation feedback case converts the cool gas into stars, and yields a 
temperature profile which follows the UTP at $r\ge 0.15r_{vir} $, but 
flattens out at smaller radii. This is consistent with the high resolution 
\textit{Chandra} observations of Vikhlinin et al. \cite{Vikhlinin04}.

\section{Comparisons and predictions for X-ray and SZE surveys}

In this section we shall compare the results of numerical hydrodynamical 
simulations with the analytic scaling laws derived in section 3, and compare 
with observational data. We will see that the X-ray temperature and the 
integrated SZE is a robust indicator of cluster mass with relatively little 
bias, while the X-ray luminosity is not because we cannot reliably simulate 
the X-ray emission from clusters. 

\subsection{Analytic and numerical comparisons}

\begin{figure}
\centerline{\includegraphics[width=4in,height=2.5in]{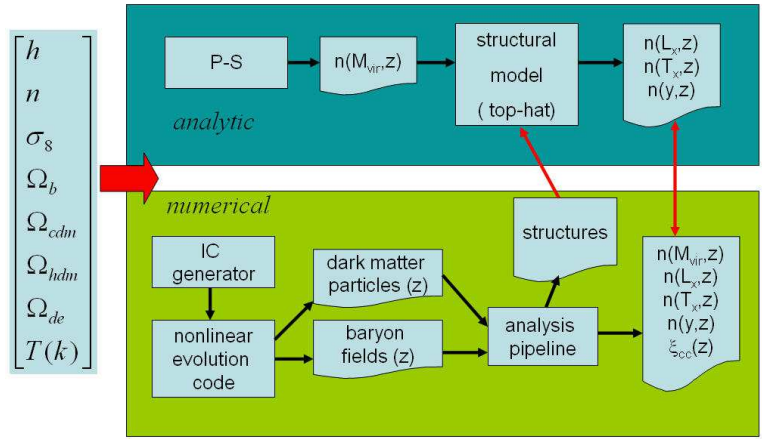}}
\caption{Comparing analytic and numerical predictions for cluster statistics.}
\label{fig11}
\end{figure}

We first ask the question how well do the simple analytic model estimates of 
cluster statistics agree with the results of numerical hydrodynamic 
simulations. This question was addressed by Bryan {\&} Norman 1998 
\cite{BN98}.
Fig. \ref{fig11} 
illustrates how the comparisons are made. For a given cosmological model 
Press-Schechter theory is used to calculate the halo mass function versus 
redshift (top rectangle). The observable quantities $n(T,z), N(L_x,z), n(Y,z)$ 
are then computed using the scaling relations presented in Section 3
for $L_x, T$ and $Y$ as a function of mass. Somewhat more work is involved deriving 
these results from numerical simulation (bottom rectangle). Initial 
conditions for the chosen cosmology are generated which specify dark matter 
and baryonic perturbations at the starting redshift. These perturbations are 
evolved use in the methods described in section 4 to z=0. The particle and 
baryonic distributions are output at specified redshifts for analysis. 
Virialized objects are located using a group-finding algorithm on the dark 
matter particles list. Two popular techniques are friends-of-friends 
\cite{Davis85} and HOP \cite{Eisenstein99}. In the friends-of-friends 
algorithm, two particles are part of the same group if their separation is 
less than some chosen value; chains of pairs then define groups. In the HOP 
algorithm, an estimate of the local density is associated with every 
particle. Each particle is linked to its densest neighbor and on to that 
particle's densest neighbor until one reaches the particle which is its own 
densest neighbor. All particles that are traced to the same such particle 
define the group. Once groups are found, centers of masses for each group 
are computed. With these centers determined, spherically averaged profiles 
of dark matter density, baryon density, temperature, etc. are computed by 
binning the 3D data into spherical shells. For each halo, the virial radius 
is determined by find the shell inside of which the mean total density (dark 
matter + baryons) equals the critical overdensity $\Delta_c$ (Section 3). Virial 
mass, X-ray luminosity, and emission weighted temperature are computed by 
numerical integration over the radial profiles of total density, X-ray 
emissivity, etc. With these quantities evaluated for each cluster in the 
sample, distribution functions are then computed. 

\subsection{Cluster temperatures}

\begin{figure}[htbp]
\includegraphics[width=3in,height=2in]{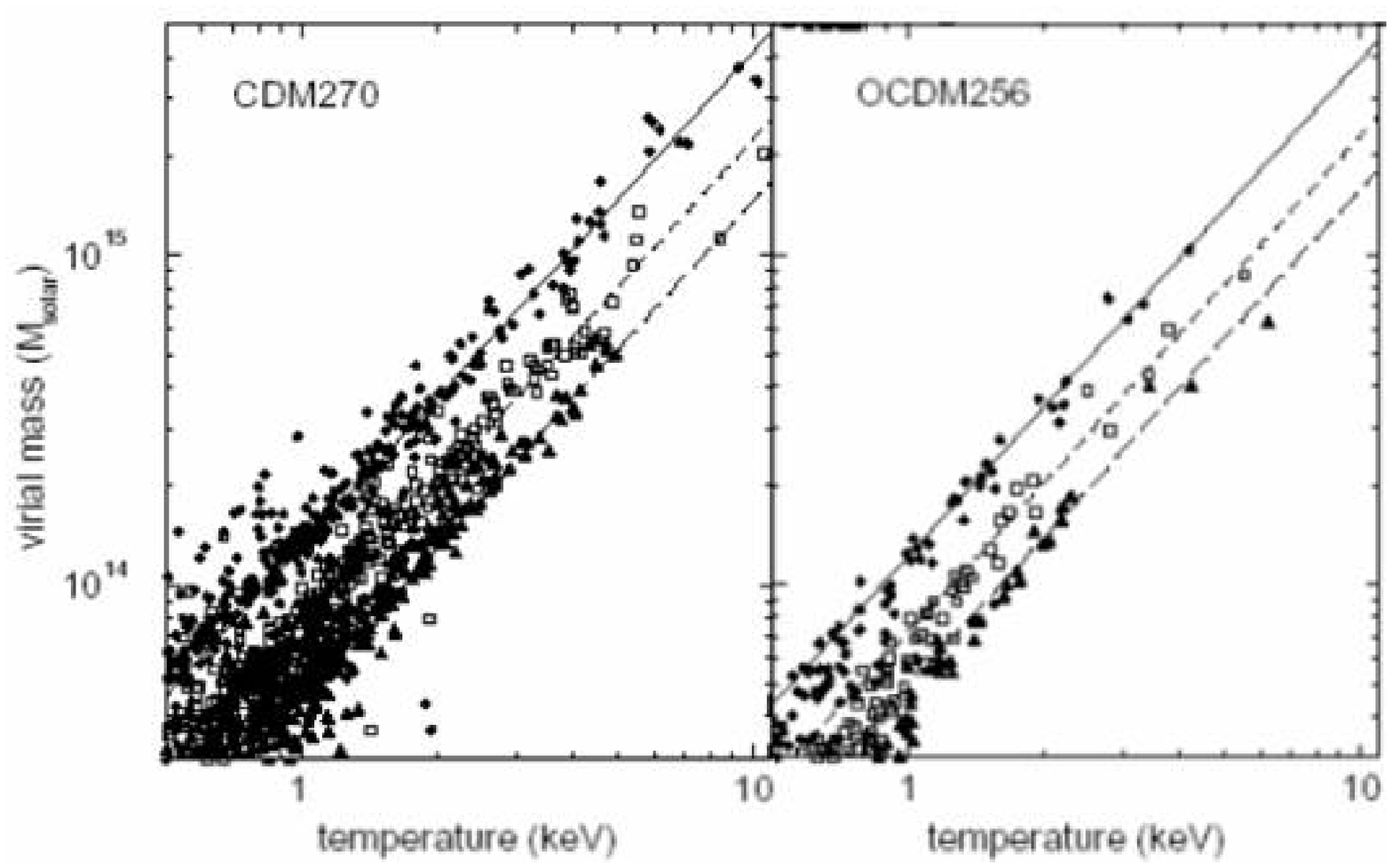}
\includegraphics[width=2.5in,height=2.5in]{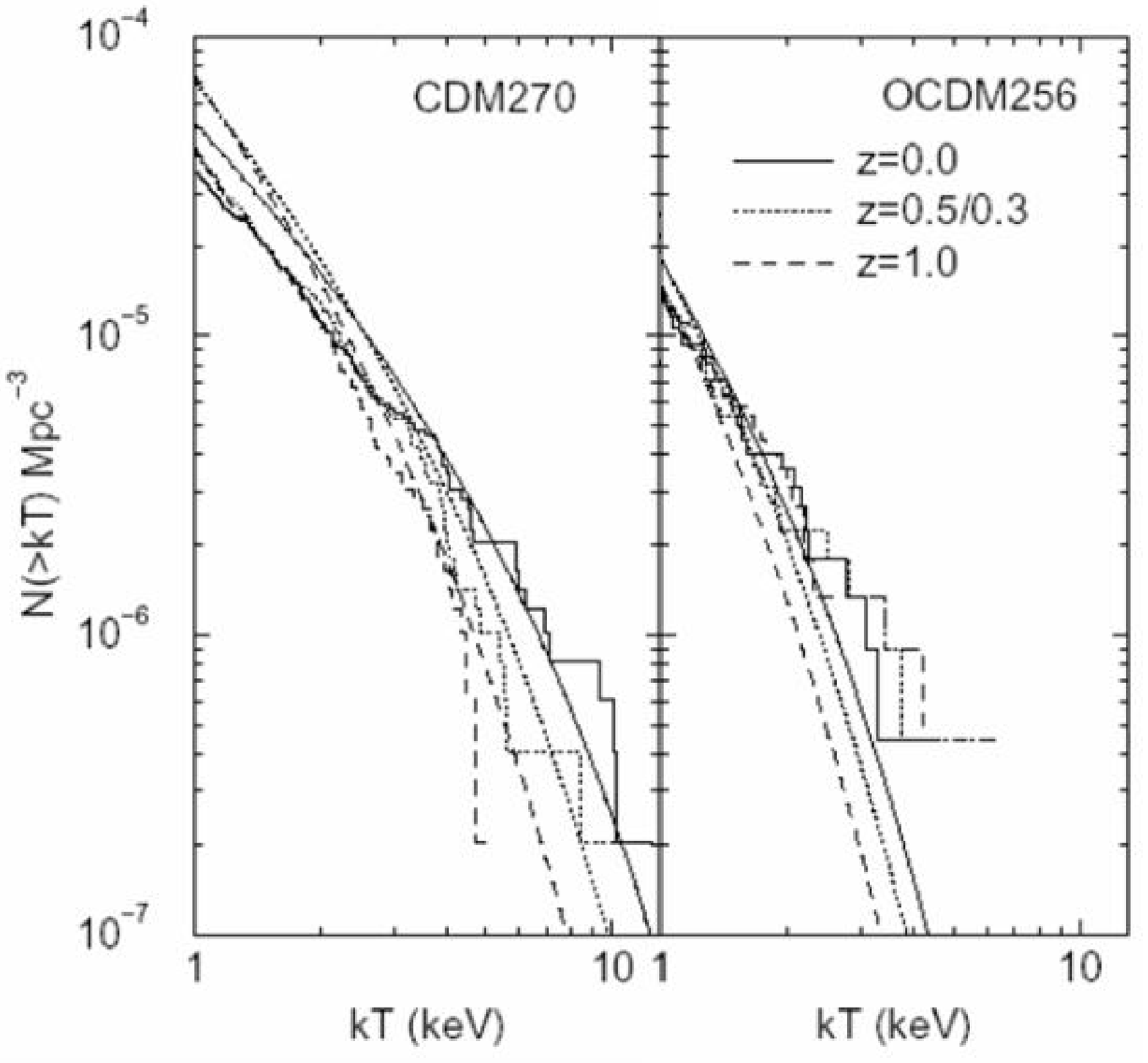}
\caption{Left: M-T scaling in a flat $\Omega _{m}$=1 universe (left) and an open $\Omega _{m}$=0.34 universe (right) for z=0, 0.5, and 1 (top to bottom). Symbols are measured values hydrodynamic simulations. Lines are the scaling relations from Eq. \ref{eq28}. with f$_{T}$=0.8 (from \cite{BN98}).
Right: Evolution of cumulative temperature distribution function for the two models shown in Fig 13 as predicted by theory (curves) and hydrodynamic simulations (histograms). The number of hot clusters evolves rapidly in the flat universe but evolves very little in the open universe.}
\label{fig12}
\end{figure}

One of the most robust predictions of numerical simulations is the 
mass-temperature relation. Fig. \ref{fig12}a shows a comparison between analytic 
scaling relations and simulations for two cosmological models at three 
epochs. The simulations were carried out on fixed Eulerian grids of size 
270$^{3}$ and 512$^{3}$ assuming the clusters are non-radiative. Good 
agreement is seen with a slight offset in normalization. Fitting Eq. \ref{eq28} to 
the data yields $f_T \approx 0.8.$ That the simulations reproduce the 
analytic scaling relations despite limited numerical resolution is a 
consequence of energy conservation, which is maintained to high accuracy by 
the numerical hydrodynamic method employed. Note that a 
cluster of a given mass is cooler at lower redshifts. 

Fig. \ref{fig12}b shows the temperature distribution function as predicted by 
simulations (histograms) and Press-Schechter theory (curves) for a 
critically closed model (SCDM) and a low density model (OCDM). Generally, 
agreement is good. Simulations underpredict the number of low temperature 
clusters due to resolution effects. The high temperature clusters are rare, 
and thus not many are found in our small box. Despite these numerical 
limitations, one sees that the number of hot clusters evolves rapidly in the 
flat universe but evolves very little in the open universe. 

\begin{figure}[htbp]
\includegraphics[width=2.5in,height=2in]{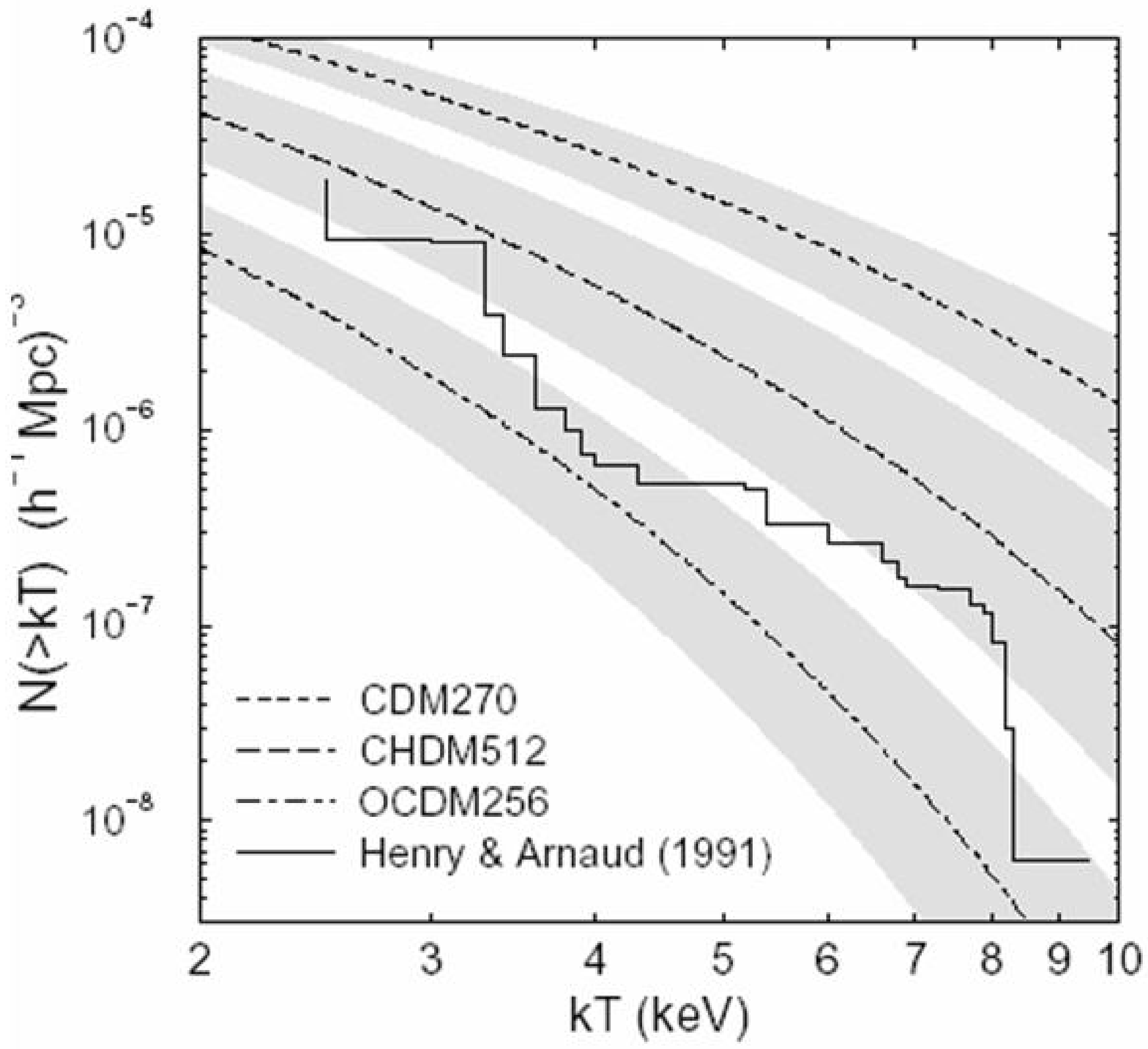}
\includegraphics[width=2.5in,height=2.5in]{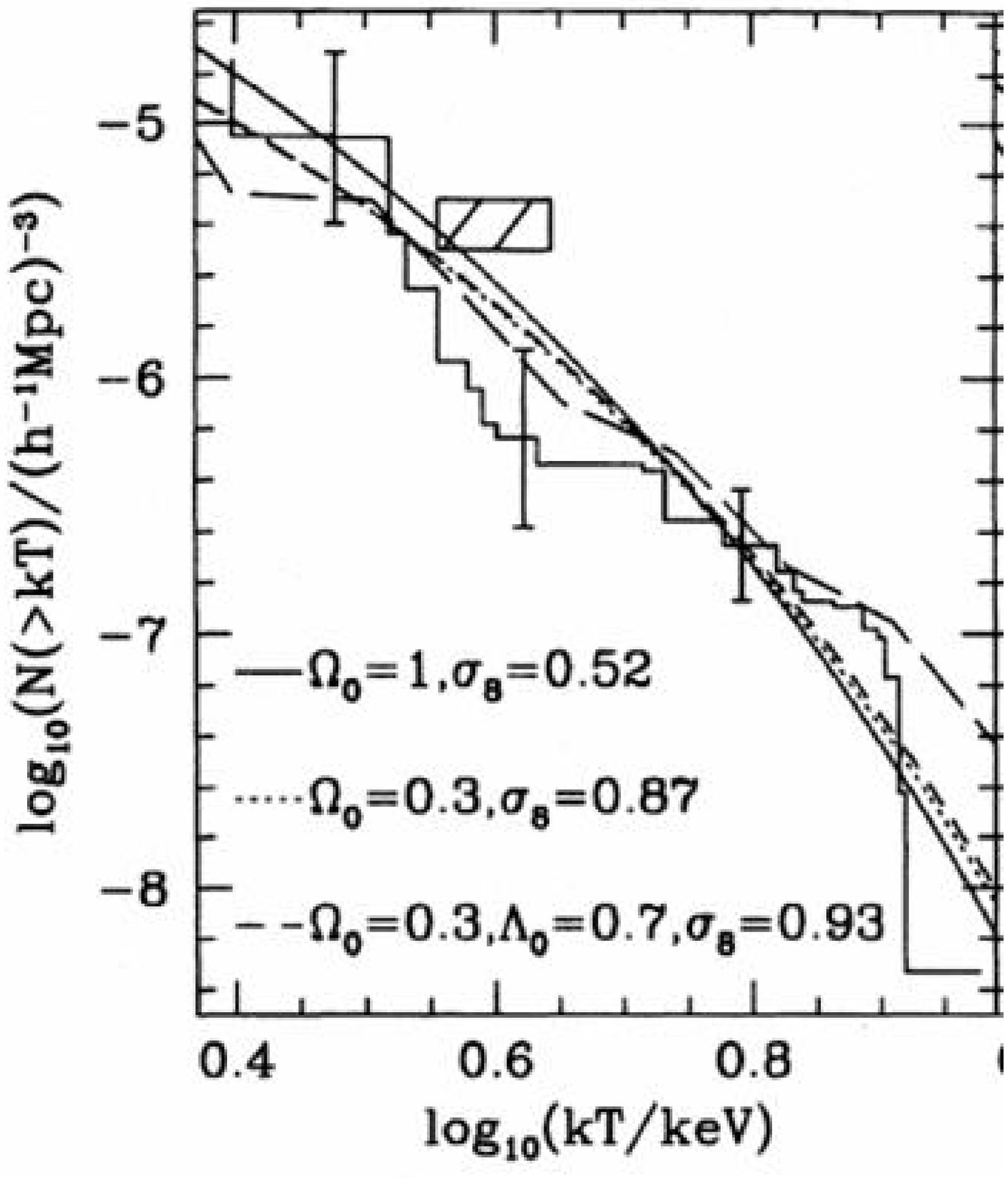}
\caption{Left: Comparison of z=0 cluster temperature function from Henry {\&} Arnaud (1991) with hydrodynamic simulations. SCDM model ($\Omega _{0}$=1, $\sigma _{8}$=1.05) is ruled out with high confidence, OCDM model ($\Omega _{0}$=0.34, $\sigma _{8}$=0.75) is marginally consistent with data. (from Bryan {\&} Norman 1998). Right:
Figure 18. Illustration of the $\Omega _{0}-\sigma _{8}$ degeneracy. Good agreement with data is found for flat, open, and $\Lambda $-dominated cosmological models with a suitable adjustment of $\sigma _{8}$. From \cite{ecf96}.}
\label{fig13}
\end{figure}

Fig. \ref{fig13}a shows the predictions of simulations compared with the 
observational data of Henry {\&} Arnaud (1991)\cite{Henry91}. 
The SCDM model is ruled out 
with high confidence, while the CHDM and OCDM models are marginally 
consistent with data. Eke, Cole {\&} Frenk (1996) \cite{ecf96} showed that with a 
suitable adjustment of $\sigma _{8}$, a critically closed, open, and 
$\Lambda $-dominated models could all reproduce the observations 
(Fig. \ref{fig13}b). 
This illustrates what is known as the $\Omega _{0}-\sigma _{8}$ 
degeneracy in cluster abundances \cite{Bahcall97}. The 
redshift evolution of cluster abundances can in principle break this 
degeneracy, however this requires large samples of high redhift clusters 
with accurately measured temperatures. So far, the samples are small. 
Temperatures are more difficult to measure than X-ray luminosities. 
Nonetheless, available data shows mild evolution of the X-ray temperature 
function, consistent with a low density universe \cite{Rosati02}.

\subsection{Cluster X-ray luminosities}

\begin{figure}
\includegraphics[width=5in,height=2.5in]{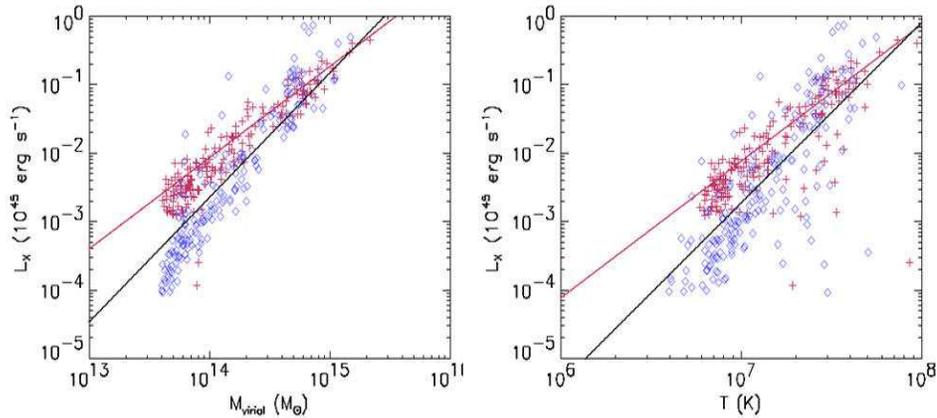}
\caption{High resolution AMR simulations of adiabatic clusters (red crosses) agree with analytic scaling predictions (red lines), but disagree with observations (black lines). Addition of radiative cooling (blue diamonds) improves agreement, but produces too many clusters with cool cores. Figures courtesy P. Motl.}
\label{fig14}
\end{figure}

The most easily measured property of an X-ray cluster is its luminosity. 
However, as we shall see, this is the most difficult quantity to predict 
using numerical simulations. This is because the integrated X-ray luminosity 
of a cluster is dominated by emission from the core region, which is 
challenging to resolve numerically, and it is affected by heating and 
cooling processes which are as yet not well understood. The advent of 
multiscale numerical simulation techniques has ameliorated the numerical 
resolution difficulties. As one can see from Fig. \ref{fig8}f, the X-ray emissivity 
peaks at about $0.1 r_{vir}$ for the adiabatic Santa Barbara cluster. SPH 
and AMR simulations can now resolve this scale with ten resolution elements 
or more in large cosmological volumes. Fig. \ref{fig14} shows the $L_x-M$ and $L_x-T$ scaling 
relation derived from our large sample of adiabatic galaxy clusters 
simulated using AMR in a $\Lambda$CDM universe. The numerical clusters are in good 
agreement with the analytic virial scaling relations $L_x \propto 
M^{4/3}$ and $L_x \propto T^2$ without resort to resolution corrections 
(cf. Bryan {\&} Norman 1998). However, the adiabatic models are in conflict 
with the observed scaling relation, which are $L_x \propto M^{1.8}$ 
 and $L_x \propto T^3$ for $T >2$ keV \cite{Rosati02}. 

\begin{figure}[htbp]
\includegraphics[width=2.5in,height=2.5in]{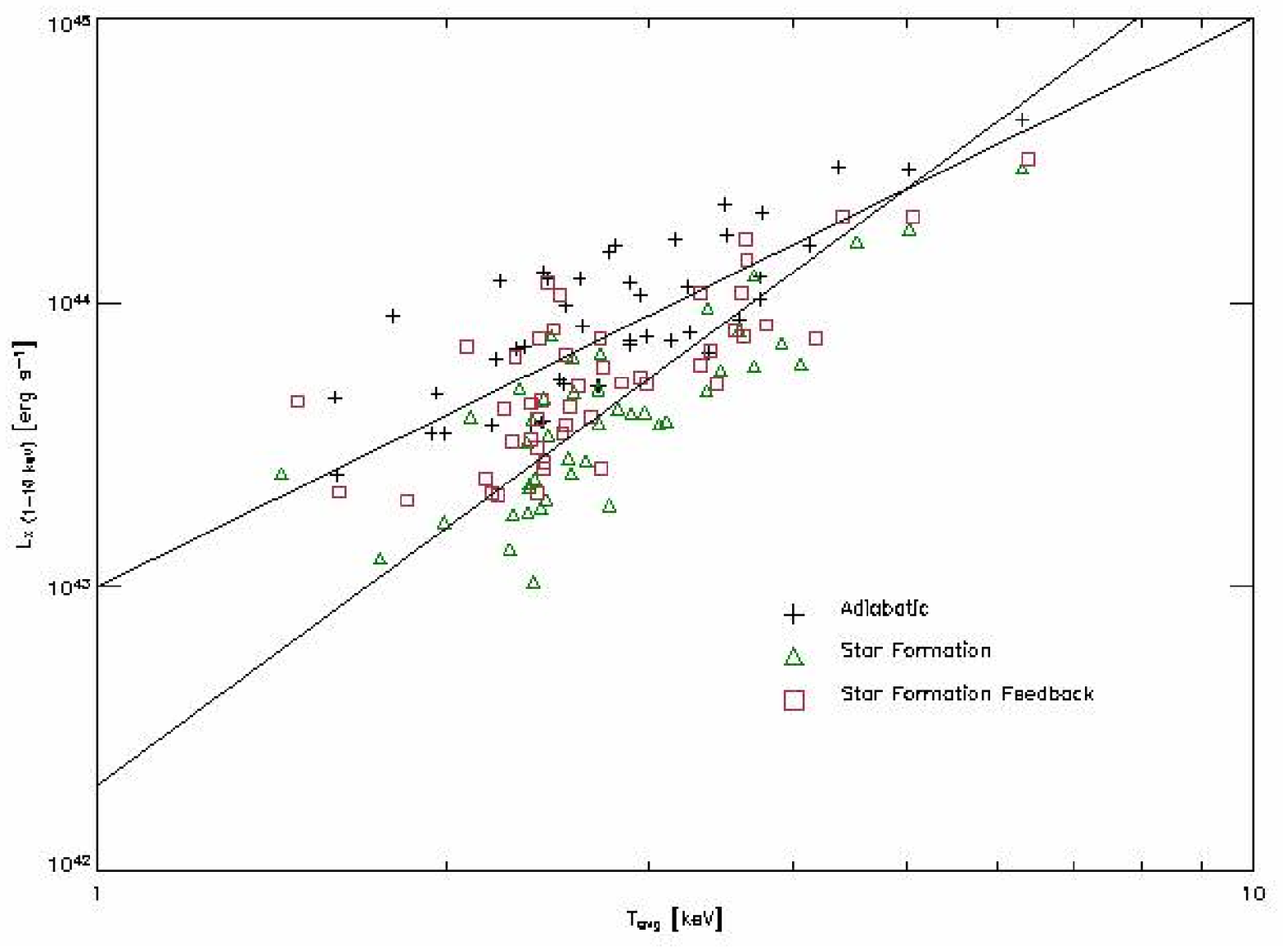}
\includegraphics[width=2.5in,height=2.5in]{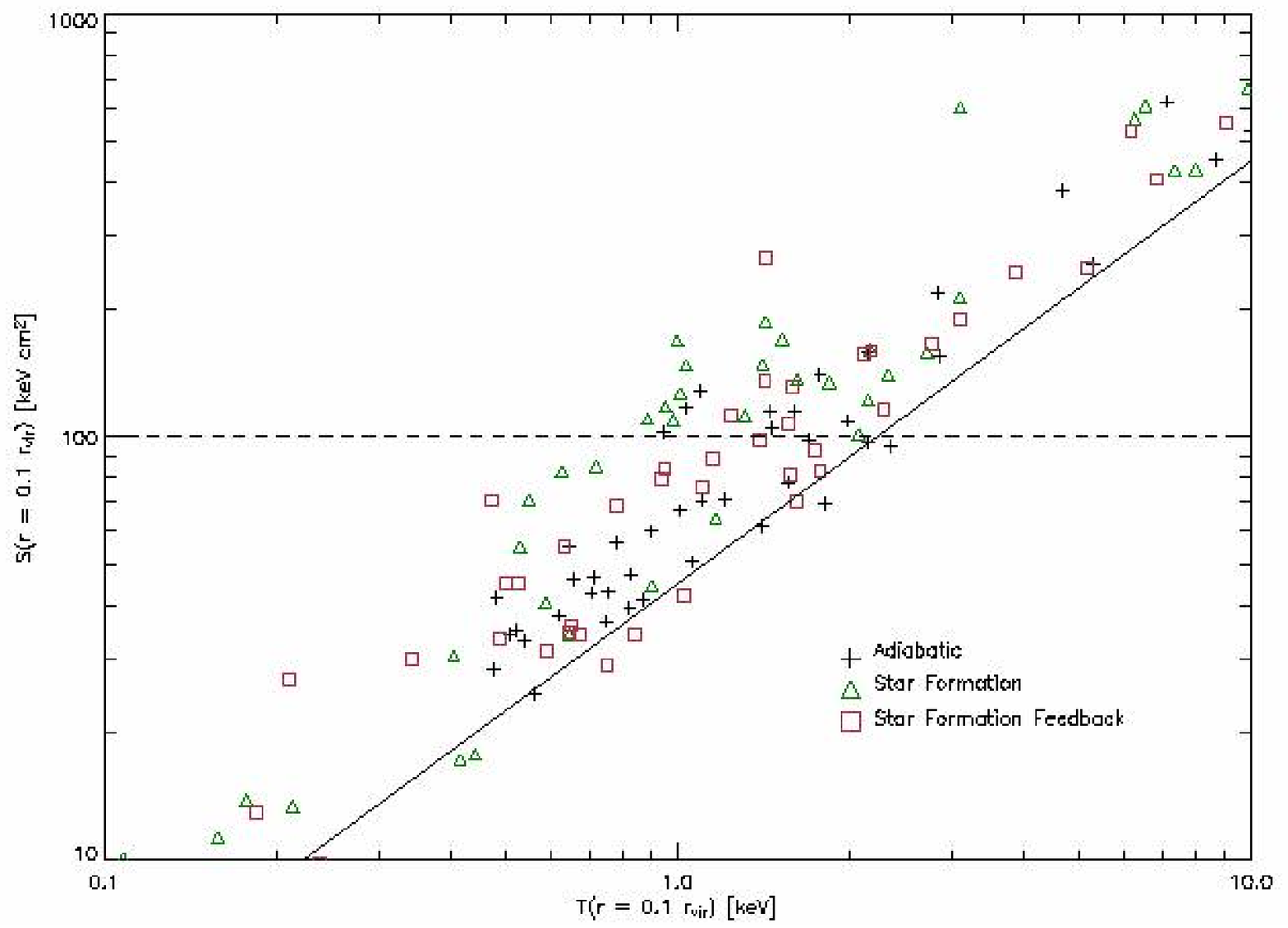}
\caption{Left: Effect of baryonic physics on the L-T relation for three 
AMR cluster samples: adiabatic (crosses), star formation (triangles), and star formation feedback (squares). Right: Central entropy versus central temperature for the cluster samples in Fig \ref{fig12}. The dashed line is the observed ``entropy floor''. Figures courtesy P. Motl.}
\label{fig15}
\end{figure}

The disagreement between the predictions of adiabatic simulations and observations 
can be taken as strong evidence of the importance of non-adiabatic processes 
in the cores of galaxy clusters. The effect of radiative cooling is shown by 
the open diamonds in Fig. \ref{fig14}. Although the $L_x-M$ and $L_x-T$ scaling steepens in the 
direction of observations, we view these models as unrealistic since every 
cluster in the sample has too much cold gas in the core, contrary to 
observations. The scaling relations for the ``star formation'' and ``star 
formation feedback'' samples are show in Fig. \ref{fig15}a. 
The conversion of cool gas 
into stars produces clusters whose temperature and X-ray surface brightness 
profiles are in better agreement with observations, and steepens the $L_x-T$ 
relation somewhat relative the to adiabatic clusters. The inclusion of 
supernova heating has a rather minor effect when compared to the magnitude 
of the change including star formation. This is best illustrated in 
Fig. \ref{fig15}b, 
which shows the scatter of central entropy versus central temperature for 
the adiabatic, star formation, and star formation feedback cluster samples. 
An analysis of a sample of clusters by Ponman et al. (1999)
\cite{Ponman99} revealed the existence of 
an ``entropy floor''. This feature has been interpreted as evidence of galaxy formation feedback which increases gas entropy. The same data has been
explained as the result of radiative cooling \cite{Bryan99,Voit00}
which locks up low entropy gas in stars where it does not contribute 
to X-ray emission. The magnitude of the entropy floor strongly suggests the 
heating explanation. The failure of star formation feedback simuations to 
exhibit the entropy floor may be due to limited mass resolution. The galaxy 
mass function is not well sampled is these simulations; indeed, only the 
central dominant galaxy and one or two of the most massive galaxies are 
present in these simulations. Perhaps higher resolution simuations will 
improve agreement. AGN heating is another source of energy input that may be 
important, especially in the cores of clusters \cite{Ruszkowski02}.
Numerical simulations incorporating these effects are in 
their infancy, and certainly not at the stage where large ensembles can be 
simulated for statistical analysis. 

\subsection{Prospects for SZE cluster surveys}

The sensitivity of X-ray luminosity to numerical resolution and baryonic 
processes motivates us to look for other more robust indicators of a 
cluster's mass. Temperature is such an indicator, however this is more 
difficult to measure than X-ray luminosity even at low redshifts. At high 
redshifts the task becomes even more difficult because of the severe $(1+z)^{-4}$ 
surface brightness dimming of the X-ray flux. In this section we explore the 
thermal SZE effect as a mass indicator based on our four catalogs of 
simulated galaxy clusters. Based on these models, we find that the 
integrated SZE $y_{500}$ is a less biased indicator of cluster mass than either 
the X-ray luminosity or temperature, and shows far less scatter than the 
central value of the SZE intensity change $y_0$. More details can be found in references
\cite{Motl05,Hallman05}

\begin{figure}[htbp]
\includegraphics[width=3in,height=2in]{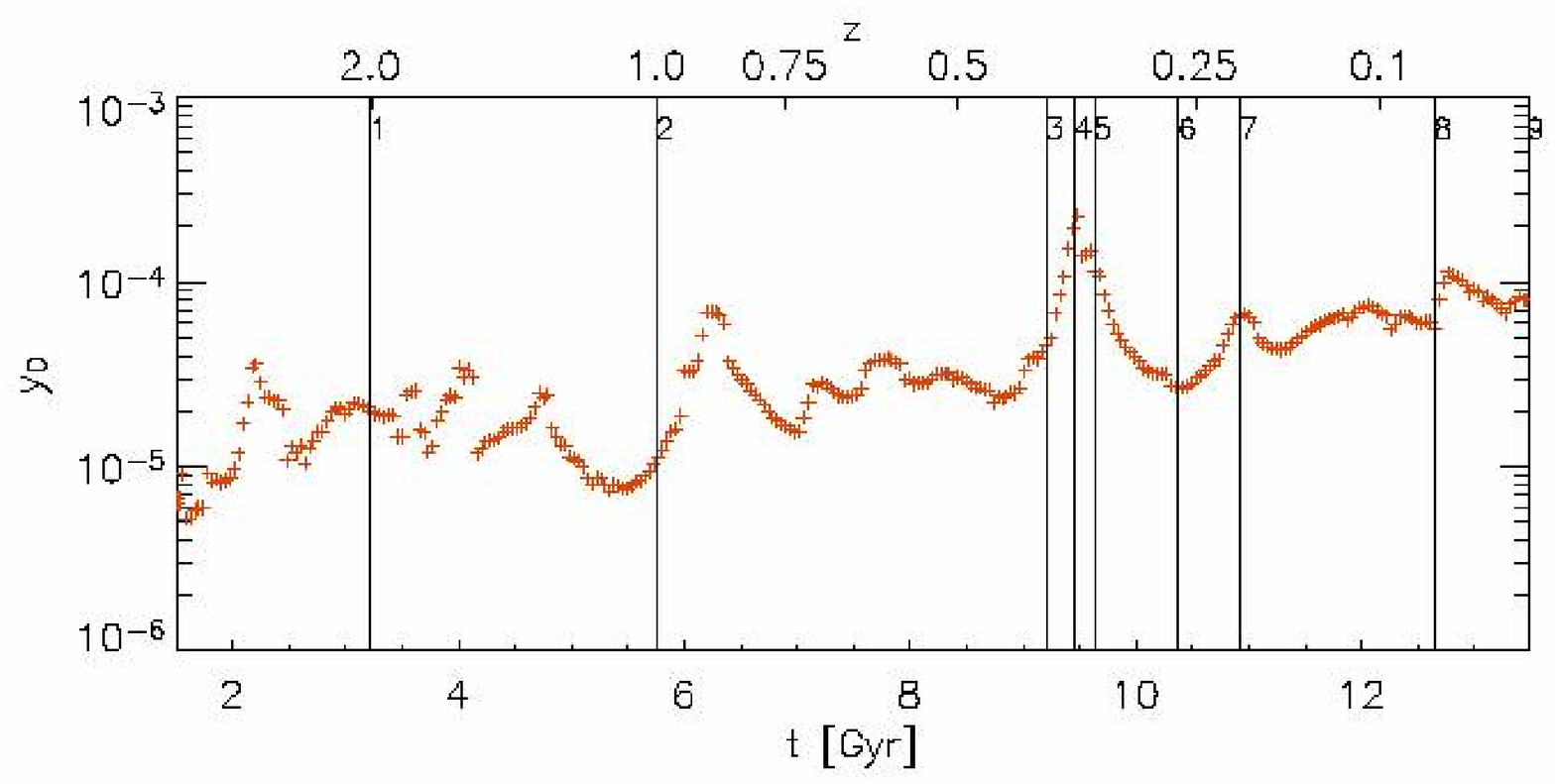}
\includegraphics[width=2in,height=2in]{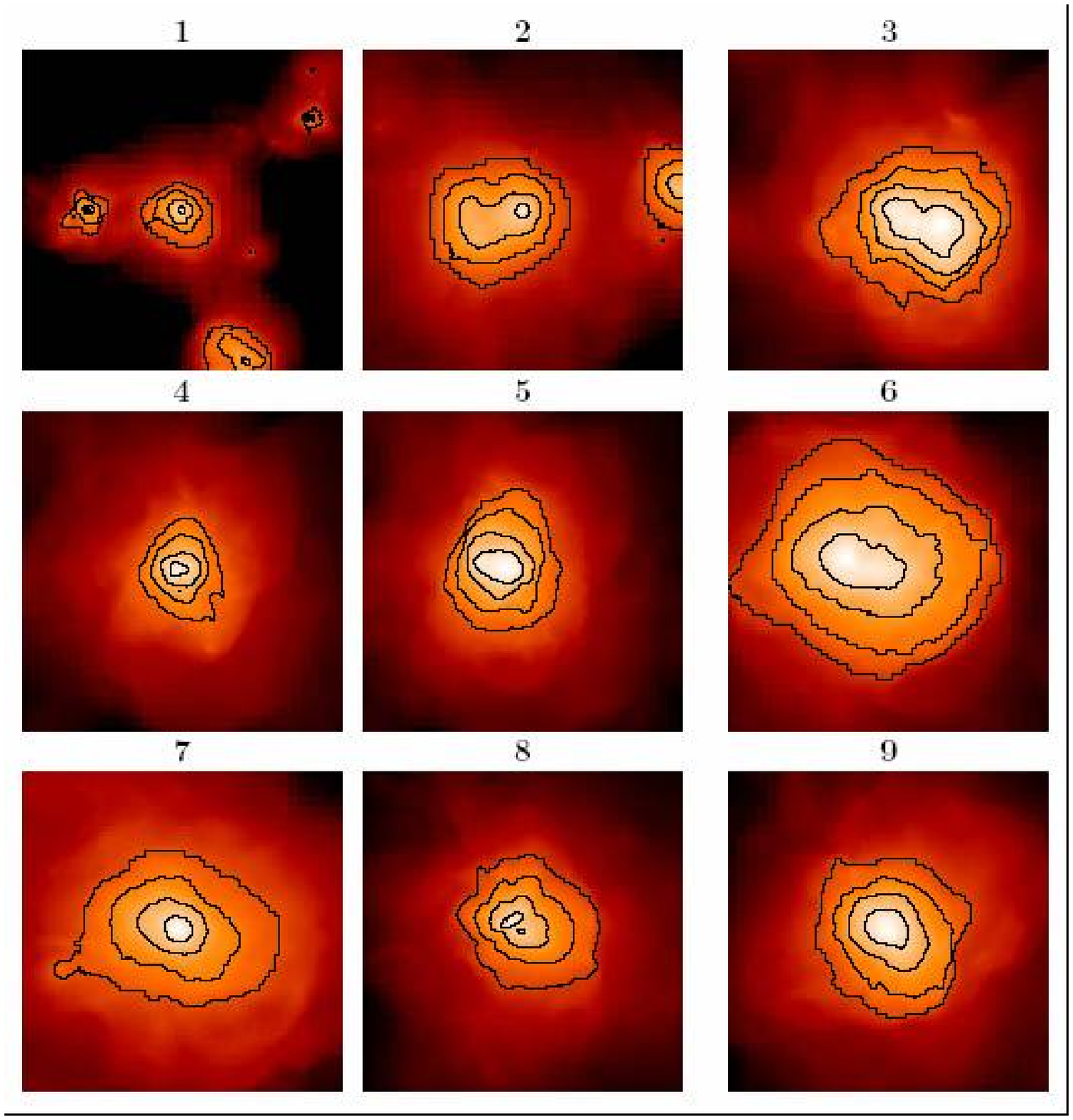}
\caption{Left: The ``lightcurve'' for the central value of the Compton parameter, 
$y_0$, obtained from tracking one particular halo from a redshift of 4 to the present epoch. Major mergers can boost $y_0$ by a factor of 10. Right:
Projected y parameter distribution of cluster at the epochs marked by vertical lines in the lightcurve. Figures courtesy P. Motl.}
\label{fig16}
\end{figure}

As has been discussed elsewhere in this volume (Rephaeli, Birkinshaw), the 
thermal SZE is an attractive cosmological probe because it is redshift 
independent. The strength of the SZE is proportional to the Compton 
parameter, y, which for non-relativistic electrons is essentially the 
integral of the gas pressure through the cluster
\begin{equation}
y=\int {\frac{k_B T}{m_e c^2}} \sigma _T n_e d\ell \propto \int {nTd\ell .} 
\end{equation}
The central value of the Compton y parameter we refer to as $y_0$. We define 
the integrated SZE $y_{500}$ as the area integral of the y parameter out to 
$r_{500}$, the radius inside of which the mean density is 500 times the critical 
density:
\begin{equation}
y_{500} =2\pi \int\limits_0^{r_{500} } {y(r)rdr.} 
\end{equation}
The detectability of a cluster is given by its SZ cross section (Section 3), 
which is essentially $y_{500} /d_A^2 \propto (1+z)^{-2}$. This is far more 
favorable redshift dependence than X-rays provide.

\begin{figure}
\centerline{\includegraphics[width=5in,height=4in]{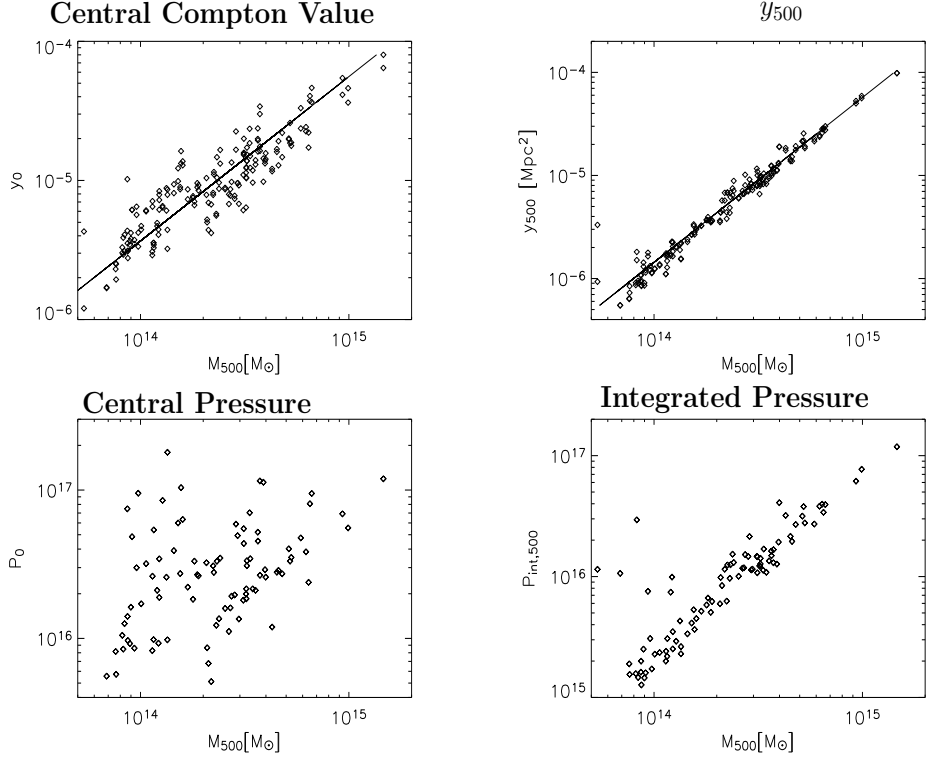}}
\caption{Upper: The scaling relations between $y_0$ and $y_{500}$ and the total cluster mass within the same radius at z=0 for the star formation with feedback cluster sample. Two randomly chosen, orthogonal projections for each cluster are  plotted as individual points and the catalog contains $\sim 100$ clusters at this epoch in the mass range $1 \times 10^{14} M_{\odot} \leq M_{200} \leq
2 \times 10^{15} M_{\odot}$. The best fit relations are plotted as solid lines. Lower: Central pressure and pressure integrated inside sphere of radius 
$r_{500}$ plotted against cluster total mass. From \cite{Motl05}.}
\label{fig17}
\end{figure}

Fig. \ref{fig16}a shows the redshift evolution of $y_0$ for the most 
massive cluster in 
our sample. As can be seen, $y_0$ exhibits a secular increase as the cluster 
potential deepens, but is boosted by up to a factor of $\sim $20(2) during 
major(minor) merger events. The duration of these events is of 
order the dynamical time $\sim $1-2 Gyr. The effect of mergers induces 
considerable scatter into scaling between $y_0$ and the enclosed mass $M_{500}$ in 
our sample of clusters at z=0 (Fig. \ref{fig17}a). By contrast, $y_{500}$ 
shows a much tighter correlation (Fig. \ref{fig17}b). The reason for this is 
illustrated in the lower two panels of Fig. \ref{fig17} where we plot the central 
value of the gas pressure $p_0$ and the volume averaged pressure $p_{500} 
=\frac{3}{4\pi r_{500}^3 }\int\limits_0^{r_{500} } {p(\vec {x})d^3\vec {x}} 
$. The central pressure exhibits large scatter due to the presence of shock 
waves induced by mergers. However, the volume averaged pressure exhibits 
relatively little scatter. This is a consequence of virial equilibrium and 
tells us that the clusters are approximately in equilibrium within $r_{500}$. 

Fitting the data to a power law of the form
\begin{equation}
y_{500} =A\left[ {\frac{M_{500} }{10^{14}M_{\odot}}} \right]^\alpha
\end{equation}
for each of our 4 catalogs, we find $\alpha \sim 1.6, \sigma_{\alpha}\sim 0.025$
for the adiabatic, star formation, and star formation feedback samples, and 
$\alpha \sim 1.7, \sigma_{\alpha}\sim 0.03$
for the radiative cooling sample. The scaling exponent 
is consistent with the findings of da Silva et al (2004) \cite{daSilva04}. 
Ignoring the 
radiative cooling only runs as unrealistic, we find that the scaling is 
relatively insensitive to baryonic physics. This is both reassuring and 
understandable in that regardless of the thermodynamics of the gas, 
hydrostatic equilibrium is maintained to a good approximation. By looking 
back through our catalogs in redshift, we find that the coefficient A is 
independent of redshift. 

\subsection{Cluster mass estimates compared}

To assess the systematic biases and relative scatter of various means of 
estimating cluster masses from X-ray and SZE data, we ``observed'' our four 
clusters samples and analyzed the resulting synthetic images in the same way 
as observations. Our goal was to find both the best cluster mass estimator 
and best method of analysis. These were defined as the combination which 
produce the least bias and smallest scatter between inferred cluster mass 
and actual (simulated) mass. Here we merely summarize our findings; for 
details the reader is referred to \cite{Hallman05}. 

Cluster masses can be obtained from X-ray and thermal SZE observations in 
several ways. The most widely used is the isothermal beta model, wherein it 
is assumed the electron number density is spherically symmetric and follows

\begin{equation}\label{eq41}
n_e (r)=n_{e0} \left[ {1+\left( {\frac{r}{r_c }} \right)^2} \right]^{-3\beta 
/2},
\end{equation}

where $n_{e0}$ is the central electron density. Approximating the gas as 
isothermal with average temperature $\langle T \rangle$ within the fitting radius, then the X-ray surface brightness is
\begin{equation}\label{eq42}
S_X (r)=S_{X0} \left[ {1+\left( {\frac{r}{r_c }} \right)^2} 
\right]^{\frac{1}{2}-3\beta }
\end{equation}
where $S_{X0} \propto n_{e0}^2 \left\langle T \right\rangle ^{\frac{1}{2}}$.
Similarly for the SZE, a beta model density distribution results in a 
projected radial distribution for the Compton y parameter
\begin{equation}\label{eq43}
y(r)=y_0 \left[ {1+\left( {\frac{r}{r_c }} \right)^2} 
\right]^{\frac{1}{2}-\frac{3\beta }{2}}
\end{equation}
where $y_{0} \propto n_{e0} \left\langle T \right\rangle.$

By fitting the observed profiles of $S_x(r)$ and $y(r)$ one obtains 
$\beta$ and $r_c$, the core radius. With $\left\langle T \right\rangle$
measured observationally, $n_{e0}$ can then be calculated. 
One then integrates Eq. \ref{eq41} to find the gas mass within the fitting 
radius $r_<$. The cluster dynamical mass is then $M_{dyn} (r_< )=M_{gas} (r_< 
)/f_b (r_< ),$ where $f_b$ is the baryon fraction which may in general be 
different from the cosmic mean $\Omega_m/\Omega_b$ depending upon the radius. Henceforth 
we will refer to mass estimates made in this way as X-ray-ISO and SZE-ISO. 

Recently is has been shown both in simulations (Loken et al. 2002, Section 4) 
and in X-ray observations (Vikhlinin et al. 2005) that clusters are not 
isothemal at large radii, but follow a universal temperature profile (UTP)
\begin{equation}\label{eq43a}
T(r)=\left\langle T \right\rangle _{500} \left[ {1+\left( {\frac{r}{\alpha 
r_{500} }} \right)^2} \right]^{-\delta }
\end{equation}
where $\langle T_{500} \rangle$ is the average temperature inside 
$r_{500}$, and $\alpha$ and $\delta$ are 
fitting parameters determined from a large sample of clusters. Improved mass 
estimates can be obtained by geometric deprojection of the X-ray and SZE 
profiles if one knows the temperature of each radial shell. This is provided 
by the UTP. For example, the X-ray surface brightness can be deprojected to 
yield the X-ray emissivity in each spherical shell (e.g., \cite{Buote00}). 
Knowing the temperature profile, once can obtain the mass in each shell. A 
similar technique can be applied to the SZE profile. By summing over shells, 
one obtains the gas mass within the fitting radius. Mass estimates obtained 
in this way we refer to as X-ray UTP and SZE-UTP. 

\begin{figure}
\centerline{\includegraphics[width=4in,height=2.5in]{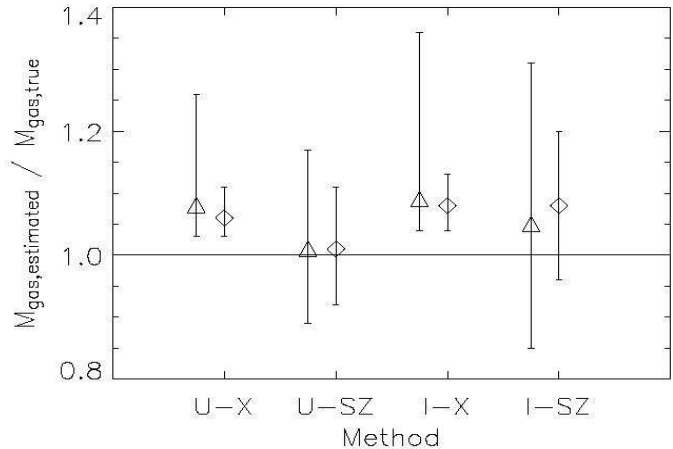}}
\caption{Comparison of median values and scatter of gas mass estimates inside
$r_{500}$ for full SFF cluster sample (triangles) and cleaned SFF sample
(diamonds) at z=0 for each of four methods: UTP-X-ray (U-X), UTP-SZE (U-SZ),
isothermal X-ray (I-X), and isothermal SZE (I-SZ) as descibed in the text.
From \cite{Hallman05}.}
\label{fig18}
\end{figure}

Fig. \ref{fig18} shows the ratio of the measured mass to the actual mass for the star 
formation feedback catalog of simulated clusters for the four methods 
described above. The triangles are the full sample, whereas the diamonds are 
for samples which have been cleaned of highly distorted clusters resulting 
from recent mergers. The error bars enclose the 80{\%} confidence range. As 
can be seen, cleaning the sample reduces the scatter considerably. Among the 
different methods, the X-ray measurements yield the smallest scatter, but 
overestimate the cluster masses by 5-10{\%}. Conversely, the SZE-UTP 
measurements yield unbiased estimates the cluster mass, with somewhat more 
scatter. As shown in \cite{Hallman05}, the scatter in the SZE estimates 
decreases as the fitting radius is increased to $r_{200}$, while no improvement 
is seen in the X-ray estimates. This is to be expected since the X-ray 
emission is heavily core-weighted, while the SZE samples larger radii. 

\subsection{Conclusions}

We have seen that galaxy clusters are sensitive cosmological probes provided 
their masses can be measured with precision. Both analytic estimates and 
numerical simulations show that the evolution of their comoving number 
density is sensitive to cosmology. With improvements in X-ray observations 
and impending large area surveys to detect clusters via the SZE, it is 
paramount to assess the accuracy to which cluster masses can be obtained 
observationally. Based on our catalogs of simulated clusters using adaptive 
mesh refinement, we find that gas masses can be measured to $\sim $10{\%} 
accuracy with 80{\%} confidence. Our study ignores instrumental or other 
observational effects. These limits in precision are a direct result of the 
deviation of the simulated clusters from simple assumptions about their 
physical and thermodynamic properties, dynamical state, and sphericity. 
Comparing a variety of methods, we find that SZE methods assuming a UTP 
produce the smallest scatter when estimating masses from a raw sample of 
clusters. Cleaning the cluster sample of obvious mergers does not improve 
the SZE estimates much, but improves the X-ray estimates substantially. As a 
practical matter, we find SZE methods are superior for mass estimation of 
large samples of clusters out to high redshift. This is particularly true if 
the cutoff radius is the virial radius, as this has the effect of smoothing 
out any boosting effects in the cluster core due to mergers.

Comparing mass estimates from our four catalogs, we find that our 
conclusions are insensitive to assumed baryonic physics, except for the 
cooling sample, which yields unrealistic-looking clusters. Mass estimates 
derived from the cooling sample are systematically high (50-100{\%}) despite 
excising the overluminous X-ray core. Reasons for this are discussed in 
detail in reference \cite{Hallman05}. 
We conclude that cool core clusters are 
poor candidates for precision mass estimation, in disagreement with previous 
studies \cite{Allen98}. 

\acknowledgments
The author is indebted to his collaborators Greg Bryan, 
Jack Burns, Eric Hallman, Chris Loken, and Patrick Motl whose results, both 
published and unpublished, are presented here. Simulations were performed at 
the National Center for Supercomputing Applications of the University of 
Illinois, Urbana-Champaign with support from NSF grants ASC-9318185, 
AST-09803137.


\begin{thebibliography}{0}
\bibitem{carlstrom02} \BY{Carlstrom, J., Holder, G, \& Reese, E.}
  \IN{Ann. Rev. Astron. Astrophys.}{40}{643}{2002}.
\bibitem{springel05} \BY{Springel, V., White, S. et al.}
  \IN{Nature}{435}{629}{2005}
\bibitem{Rosati02} \BY{Rosati, P., Borgani, S., \& Norman, C.}
  \IN{Ann. Rev. Astron. Astrophys.}{40}{539}{2002}.
\bibitem{dodelson03} \BY{Dodelson, S.} 
  \TITLE{Modern Cosmology},~(Academic Press, Amsterdam), 2003.
\bibitem{Perlmutter03} \BY{Perlmutter, S.}
  \IN{Physics Today}{April 2003}{53}
\bibitem{spergel03} \BY{Spergel, D. et al.}
  \IN{ApJS}{148}{175}{2003}.
\bibitem{bops99} 
  \BY{Bahcall, N. A.; Ostriker, J. P.; Perlmutter, S.; Steinhardt, P. J.}
  \IN{Science}{284}{1481}{1999}
\bibitem{tegmark03} \BY{Tegmark, M. et al.}
  \IN{ApJ}{606}{702}{2004}.
\bibitem{wef93} \BY {White, S, Efstathiou, G., \& Frenk, C.}
  \IN{MNRAS}{262}{1023}, 1993.
\bibitem{kolbturner90} \BY{Kolb, E. \& Turner, M.}
  \TITLE{The Early Universe},~(Addison-Wesley, Redwood City, CA), 1990.
\bibitem{komatsu03} \BY{Komatsu, E. et al.}
  \IN{ApJS}{148}{119}{2003}.
\bibitem{pad93} \BY{Padmanabhan, T.}
  \TITLE{Structure Formation in the Universe},~(Cambridge University Press, Cambridge), 1994.
\bibitem{BN98} \BY{Bryan, G. \& Norman, M.}
  \IN{ApJ}{495}{80}{1998}.
\bibitem{kaiser86} \BY{Kaiser, N.}
  \IN{MNRAS}{222}{323}{1986}.
\bibitem{bt87} \BY{Binney, J. \& Tremaine, S.}
  \TITLE{Galactic Dynamics},~(Princeton University Press, Princeton, USA), 1987.
\bibitem{ps74} \BY{Press, W. \& Schechter, S.}
  \IN{ApJ}{187}{425}{1974}.
\bibitem{white94} \BY{White, S. D. M.}
  \TITLE{Cosmology and Large Scale Structure: Proceedings of Les Houches Summer School}, R. Schaeffer et al., editors, (Elsevier, Amsterdam), 1996.
\bibitem{ecf96} \BY{Eke, V., Cole, S. \& Frenk, C.}
  \IN{MNRAS}{281}{703}
\bibitem{norman03} \BY{Norman, M. L.}
  \TITLE{Matter and Energy in Clusters of Galaxies, ASP Conference Series Vol. 301},
  S. Boyer \& C.-Y. Hwang, eds., (Astronomical Society of the Pacific, San         Francisco), p. 419, 2003.
\bibitem{Frenk99} \BY{Frenk, C. et al.}
  \IN{ApJ}{525}{554}{1999}
\bibitem{KWH96} \BY{Katz, N., Weinberg, D. \& Hernquist, L.}
  \IN{ApJS}{105}{19}{1996}
\bibitem{syw01} \BY{Springel, V., Yoshida, N., \& White, S.}
  \IN{NewA}{6}{79}{2001}
\bibitem{bn97} \BY{Bryan \& Norman, M.}
  \TITLE {Computational Astrophysics; 12th Kingston Meeting on Theoretical Astrophysics}, D. A. Clarke and M.~Fall, editors, ASP
  Conference Series \# 123, 1997.
\bibitem{Kravtsov97} \BY{Kravtsov, A., Klypin, A., \& Kokhlov, A.}
  \IN{ApJS}{111}{73}{1997}
\bibitem{Teyssier02} \BY{Teyssier, R.}.
  \IN{Astron. Astrophys.}{385}{337}{2002}
\bibitem{OShea04} \BY{O'Shea, B. et al.}
  \TITLE{Adaptive Mesh Refinement--Theory and Applications}, T. Plewa 
  et al., eds., Springer Lecture Notes in Computational Science \& Engineering,
  (Springer, Berlin), 2005.
\bibitem{springel00} \BY{Springel, V. et al.}
  \IN{MNRAS}{328}{726}{2001}
\bibitem{Anninos97} \BY{Anninos, P. et al.}
  \IN{NewA}{2}{209}{1997}
\bibitem{Ponman99} \BY{Ponman, T., Cannon, D. \& Navarro, J.}
  \IN{Nature}{397}{135}{1999}
\bibitem{Efstathiou81} \BY{Efstathiou, G. et al.}
  \IN{ApJS}{57}{241}{1985}
\bibitem{Couchman91} \BY{Couchman, H.}
  \IN{ApJL}{368}{L23}{1991}
\bibitem{BarnesHut86} \BY{Barnes, J. \& Hut, P.}
  \IN{Nature}{324}{446}{1986}
\bibitem{WarrenSalmon94} \BY{Warren, M. \& Salmon, J.}
  \IN{Comp. Phys. Comm.}{87}{266}{1995}
\bibitem{Xu99} \BY{Xu G.} \IN {ApJS}{98}{355}{1995}
\bibitem{Evrard88} \BY{Evrard, A.} \IN{MNRAS}{235}{911}{1988}.
\bibitem{Kravtsov03} \BY{Kravtsov, A., Klypin, A. \& Hoffman, Y.}
  \IN{ApJ}{571}{563}{2002}
\bibitem{Kang94} \BY{Kang, H. et al.}
  \IN{ApJ}{428}{1}{1994}
\bibitem{Bryan94} \BY{Bryan, G. et al.}
  \IN{ApJ}{428}{405}{1994}.
\bibitem{Borgani04} \BY{Borgani, S. et al.} \IN{MNRAS}{348}{1078}{2004}
\bibitem{Berger89} \BY{Berger, M. \& Colella, P.}
  \IN{J. Comp. Phys.}{82}{64}{1989}
\bibitem{Hockney88} \BY{R.~Hockney and J.~Eastwood}
  \TITLE{Computer Simulation Using Particles}, (McGraw Hill, New York), 1988.
\bibitem{Collela84} \BY{P.~Colella and P.~R. Woodward}
  \IN{J. Comp. Physics}{54}{174}{1984}
\bibitem{Anninos94} \BY{W.~Y. Anninos \& M.~L. Norman}
 \IN{ApJ}{429}{434}{1994}
\bibitem{Cen92} \BY{Cen, R. \& Ostriker, J.}
 \IN{ApJ}{417}{404}{1993}
\bibitem{NFW96} \BY{Navarro, J., Frenk, C. \& White, S.}
  \IN{ApJ}{462}{563}{1996}
\bibitem{Motl05} \BY{Motl, P. et al.} \IN{ApJL}{623}{L63}{2005}
\bibitem{Loken02} \BY{Loken, C. et al.} \IN{ApJ}{579}{571}{2002}
\bibitem{Motl04} \BY{Motl, P. et al.}
  \IN{ApJ}{606}{635}{2004}.
\bibitem{Hallman05} \BY{Hallman, E. et al.}
  \TITLE{preprint}, astro-ph/0509460
\bibitem{Vikhlinin04} \BY{Vikhlinin, A. et al.}
  \IN{ApJ}{628}{655}{2005}
\bibitem{Ascasibar03} \BY{Ascasibar, Y. et al.}
  \IN{MNRAS}{346}{731}{2003}
\bibitem{Davis85} \BY{Davis, M. et al.} \IN{ApJ}{292}{371}{1985}
\bibitem{Eisenstein99} \BY{Eisenstein, D. \& Hut, P.}
  \IN{ApJ}{498}{137}{1998}
\bibitem{Henry91} \BY{Henry, J. P. \& Arnaud, K.}
  \IN{ApJ}{372}{410}{1991}
\bibitem{Bahcall97} \BY{Bahcall, N., Fan, X. \& Cen, R.}
  \IN{ApJ}{485}{L53}{1997}
\bibitem{Bryan99} \BY{Bryan, G.} \IN{ApJ}{544}{L1}{2000}
\bibitem{Voit00} \BY{Voit, M. \& Bryan, G.} \IN{ApJ}{551}{L139}{2001}
\bibitem{Ruszkowski02} \BY{Ruszkowski, M., Bruggen, M. \& Begelman, M.}
  \IN{611}{158}{2004}
\bibitem{daSilva04} \BY{da Silva, A. et al.} \IN{MNRAS}{348}{1401}{2004}
\bibitem{Buote00} \BY{Buote, D. A.} \IN{ApJ}{539}{172}{2000}
\bibitem{Allen98} \BY{Allen, S. \& Fabian, A.}
  \IN{MNRAS}{297}{L57}{1998}
\bibitem{Markevitch98} \BY{Markevitch, M. et al.}
  \IN{ApJ}{503}{77}{1998}
\bibitem{deGrandi02} \BY{De Grandi, S. \& Molendi, S.}
  \IN{ApJ}{567}{163}{2002}
\bibitem{Fabian94} \BY{Fabian, A. C.}
  \IN{ARAA}{32}{277}{1994}
\end{thebibliography}
\end{document}